\documentclass[12pt]{iopart}
\usepackage[latin1]{inputenc}
\usepackage{enumerate}

\usepackage{graphicx}
\usepackage{pifont}
\usepackage{cite}   % must appear before hyperref
\usepackage{hyperref}

\renewcommand{\exp}{\mathrm{exp}}
\newcommand{\A}{\mathrm{A}}
\newcommand{\B}{\mathrm{B}}
\newcommand{\C}{\mathrm{C}}
\newcommand{\D}{\mathrm{D}}
\newcommand{\arr}{\mathrm{arr}}
\newcommand{\dep}{\mathrm{dep}}
\newcommand{\pen}{\mathrm{pen}}
\newcommand{\crit}{\mathrm{crit}}
\newcommand{\res}{\mathrm{res}}
\renewcommand{\max}{\mathrm{max}}

\newcommand{\COne}{\ding{172}\;}
\newcommand{\CTwo}{\ding{173}\;}
\newcommand{\CThree}{\ding{174}\;}
\newcommand{\CFour}{\ding{175}\;}

\newcommand{\figsrc}{figs/}
\newcommand{\figwidth}{110mm}

\makeatletter
\renewenvironment{figure}[1][]{\@float{figure} \centering}{\end@float}
\makeatother

\begin{document}

\title{Self-Control of Traffic Lights and Vehicle Flows in Urban Road Networks}

\author{Stefan L\"{a}mmer$^1$ and Dirk Helbing$^{2,3}$}

 \address{$^1$
    Dresden University of Technology,\\
    Faculty of Transportation and Traffic Sciences ``Friedrich List'',\\
    A.-Schubert-Str. 23, 01062 Dresden, Germany}

 \address{$^2$
    ETH Zurich, Swiss Federal Institute of Technology,\\
    UNO D 11, Universit\"atstr. 41, 8092 Zurich, Switzerland}

 \address{$^3$
    Collegium Budapest~-- Institute for Advanced Study,\\
    Szenth\'{a}roms\'{a}g utca 2, 1014 Budapest, Hungary}
% \ead{dhelbing@ethz.ch}

\begin{abstract}

Based on fluid-dynamic and many-particle (car-following) simulations of traffic flows in
(urban) networks, we study the problem of coordinating incompatible traffic flows at
intersections. Inspired by the observation of self-organized oscillations of pedestrian
flows at bottlenecks [D.~Helbing and P.~Moln\'{a}r, {\em Phys. {R}ev. {E}} {\bf 51}
(1995) 4282--4286], we propose a self-organization approach to traffic light control. The
problem can be treated as multi-agent problem with interactions between vehicles and
traffic lights. Specifically, our approach assumes a priority-based control of traffic
lights by the vehicle flows themselves, taking into account short-sighted anticipation of
vehicle flows and platoons. The considered local interactions lead to emergent
coordination patterns such as ``green waves'' and achieve an efficient, decentralized
traffic light control. While the proposed self-control adapts flexibly to local flow
conditions and often leads to non-cyclical switching patterns with changing service
sequences of different traffic flows, an almost periodic service may evolve under certain
conditions and suggests the existence of a spontaneous synchronization of traffic lights
despite the varying delays due to variable vehicle queues and travel times. The
self-organized traffic light control is based on an optimization and a stabilization
rule, each of which performs poorly at high utilizations of the road network, while their
proper combination reaches a superior performance. The result is a considerable reduction
not only in the average travel times, but also of their variation. Similar control
approaches could be applied to the coordination of logistic and production processes.

\vspace{2pc}

\end{abstract}


\maketitle

% ------------------------------------------------------------------------------------
% ------------------------------------------------------------------------------------
% ------------------------------------------------------------------------------------
\section{Introduction}
\label{sec:Introduction}

Within the USA alone, the cost of congestion per year is estimated to be 63.1
billion~US~\$, caused by 3.7 billion hours of delays and 8.7 billion liters of ``wasted''
fuel \cite{Schrank2005}. The urgency to reduce CO$_2$ emissions and fuel consumption, and
the excessive, unpredictable travel times during traffic congestion, however, call for
more flexible and efficient control approaches. The grand challenge of travel time
minimization is the coordination of vehicle flows and, in particular, of traffic lights.

Traffic systems are a prominent example of non-equilibrium systems and have been studied
extensively in the field of statistical physics
\cite{Helbing1997a,Chowdhury2000,Schadschneider2002}. Much attention was devoted to the
study of self-organized phenomena in driven many-particle systems \cite{Helbing2001} such
as pedestrian flows \cite{Helbing1995,Helbing2006c} or traffic flows on highways
\cite{Nagatani2002,Kerner2004}. In order to explain phenomena like the emergence of
traffic jams \cite{Nagatani1999,Nishinari2003} or stop-and-go waves
\cite{Kerner2002,Banks1999,Lighthill1955}, a huge variety of different traffic flow
models have been proposed, e.g. follow-the-leader models \cite{Treiber2000} or
fluid-dynamic traffic models in both, discrete \cite{Nagel1992} and continuous
\cite{Daganzo1994a,Helbing2003a} space. More recently, a research focus was put on
network traffic, which required to extend one-dimensional traffic models in order to cope
with situations, where traffic flows merge or intersect
\cite{Daganzo1995a,Esser1997,Helbing2006c,Helbing2005,Helbing2005a,Helbing2007}. These
models can explain how jam fronts propagate backwards over network nodes
\cite{Wastavino2007,Gugat2005}, which might eventually result in cascading break-downs of
network flows \cite{Daganzo2007,Zheng2007,Simonsen2007}.

One major challenge in this connection is the optimization of traffic lights in urban
road networks \cite{Helbing2007}, especially the coordination among them. A typical goal
is to find optimal cycle times \cite{Brockfeld2001,Fouladvand2001} and to study the
corresponding statio-temporal patterns of traffic flow
\cite{Chowdhury1999,Huang2003,Sasaki2003}. It is agreed, however, that a further
improvement of the traffic flow requires to apply more flexible strategies than
fixed-time controls \cite{Sekiyama2001,Lammer2007,Fouladvand2004,Barlovic2004}.
Gershenson \cite{Gershenson2005}, for example, showed for a regular network with periodic
boundary conditions that his control strategy synchronizes traffic lights even without
explicit communication between them. Lämmer et al. \cite{Lammer2006} proposed to
represent the traffic lights by locally coupled phase oscillators, whose frequencies
adapt to the minimum cycle of all nodes in the network. Further algorithms perform
parameter adaptations by means of neural networks \cite{Nakatsuji1995,Ledoux1997},
genetic reinforcement learning \cite{Mikami1994}, fuzzy logic \cite{Chiu1993,Trabia1999},
or swarm algorithms \cite{Hoar2002}.

The optimization of intersecting network flows has also been studied in the domain of
production \cite{Perkins1994,Chase1992,Burgess1997,Righter2002} and control theory
\cite{Savkin2002,Lefeber2004,Lan2006,Eekelen2007a}. De Schutter and de Moor
\cite{Schutter1999,Schutter2002} proposed a solution approach for finding optimal
switching schedules for an isolated intersection with constant arrival rates. For
networks of more than one node, Lefeber and Rooda~\cite{Lefeber2006} could derive a
state-feedback controller from a given desired global network behaviour. Besides
optimality, control theorists particularly addressed the issue of stability of
decentralized control strategies \cite{Perkins1989,Kumar1995,Burgess1997}. Whereas
so-called clearing policies (see \ref{sec:dynamicInstabilities}), for example, stabilize
single nodes in isolation, they might cause instabilities in networks with bidirectional
flows \cite{Kumar1990,Humes1994,Reiman1998,Dai2004}. Control strategies based on periodic
switching sequences, e.g. the classical fixed-time traffic light control, have been shown
to be both stable and controllable under certain conditions \cite{Savkin1998,Savkin2002}.

In this paper, we propose a decentralized control algorithm, which is based on short-term
traffic forecasts \cite{Lammer2007a} and enables coordination among neighboring traffic
lights. Rather than optimizing globally for {\em assumed} flow conditions that are never
met exactly, we look for a heuristics that most of the time comes close to optimal
operation, given the {\em actual} traffic situation. Assuming that it would be possible
to adjust traffic regulations accordingly, we will drop the condition of periodic
operation to allow for more flexible adjustment to varying traffic flows.

The fact that varying traffic flows influence the respective traffic lights ahead, which
in turn influence the traffic flows, makes it \emph{im}possible to predict the evolution
of the system over longer time horizons. This makes large-scale coordination among
traffic lights difficult. It is known, however, that local non-linear interactions can,
under certain conditions, lead to system-wide spatio-temporal patterns of motion
\cite{Wiggins2003}. Therefore, our control concept pursues a local self-organization
approach. The particular scientific challenge is that such a decentralized
``self-control'' must be able to cope with (1) real-time optimization, (2) feedback loops
due to the mutual interaction between the traffic lights via the traffic flows, and (3)
very limited prognosis horizons.

Our paper is organized as follows: In the next section, we introduce a fluid-dynamic
model for the traffic flow in urban road networks. This model allows us to anticipate the
effects of switching traffic lights (see Sec.~\ref{sec:Anticipation}). In
Sec.~\ref{sec:TrafficLightControl}, we explain our concept of self-control of traffic
lights. The underlying principle is inspired by the self-organization of opposite
pedestrian flows, which is driven by the pressure differences between the waiting crowds.
We generalize this observation in Sec.~\ref{sec:Prio} to define priorities of arriving
traffic flows. In Secs.~\ref{sec:Stab} and~\ref{sec:Combined}, the prioritization
strategy is supplemented by a stabilization strategy. Simulation studies are presented in
Sec.~\ref{sec:Simulation} and demonstrate the superior performance of our decentralized
concept of self-control.

% ------------------------------------------------------------------------------------
% ------------------------------------------------------------------------------------
% ------------------------------------------------------------------------------------
\section{Network flow model}
\label{sec:FlowModel}

An urban road network can be composed of links (road sections of homogeneous capacity)
and nodes (intersections, merges, and diverges) defining their connection. The following
sections summarize a fluid-dynamic model describing the traffic dynamics on the
constituents of a road network.

% ------------------------------------------------------------------------------------
% ------------------------------------------------------------------------------------
\subsection{Traffic dynamics on road sections resulting from the continuity equation}

Let us consider a homogenous road section $i$ with constant, i.e. time-invariant length
$L_i$, speed limit $V_i$, and saturation flow $Q_i^\max$. The traffic dynamics on the
road section can be characterized by the arrival rate $Q_i^\arr(t) \leq Q_i^\max$ and the
departure rate $Q_i^\dep(t) \leq Q_i^\max$. These quantities represent the numbers of
vehicles per unit time entering or leaving the road section over all its lanes.

The flow of traffic along an urban road section (in contrast to freeway sections
\cite{Kerner2002}) is sufficiently well represented by Lighthill and Whitham's fluid
dynamic traffic model \cite{Lighthill1955}. It describes the spatio-temporal dynamics of
congestion fronts based on the continuity equation for vehicle conservation, plus a
flow-density relationship known as ``fundamental diagram''. If we neglect net effects of
overtaking and approximate the fundamental diagram by a triangular shape, this implies
two distinct characteristic speeds: While perturbations of free traffic flow propagate
downstream at the speed $V_i$, in congested traffic the downstream jam front and
perturbations propagate upstream with a characteristic speed of about -15 km/h
\cite{Helbing2001}. These fundamental relations also allow to derive explicit expressions
for the motion of the \emph{upstream} jam front, where vehicles brake and enter the
congested area of the road section, as well as for the related travel times
\cite{Helbing2005,Helbing2003a}.

An integration over space results in an effective queueing-theoretical traffic model
based on coupled delay-differential equations \cite{Helbing2007}. It can be summarized as
follows: In free traffic, ideally, the cumulated number $N_i^\exp(t)$ of vehicles
expected to reach the downstream end of road section $i$ until time $t$ is given by
\begin{equation}
    \label{eq:Nexp}
    N_i^\exp(t) = \int_{-\infty}^t Q_i^\arr \left( t' - L_i / V_i \right) dt'    \,,
\end{equation}
where the time shift $L_i/V_i$ corresponds to the travel time to pass link $i$ in free
traffic. In case of congestion, however, the number of vehicles that have actually left
the road section at its downstream end is given by the integral of the departure rate:
\begin{equation}
    \label{eq:Ndep}
    N_i^\dep(t) = \int_{-\infty}^t Q_i^\dep(t') dt' \leq N_i^\exp(t)    \,,
\end{equation}
Thus, the difference between $N_i^\exp(t)$ and $N_i^\dep(t)$ directly corresponds to the
number of delayed vehicles, which will be refered to as the queue length $n_i(t)$.
Consequently, the total waiting time $w_i(t)$ of all vehicles on road section $i$ until
time $t$ increases at the rate
\begin{equation}
    \label{eq:n}
    d w_i / dt = n_i(t) = N_i^\exp(t) - N_i^\dep(t)\,.
\end{equation}
It is important to note that even though $n_i(t)$ does not explicitly account for the
spatial location of congestion on link $i$, it fully captures the corresponding
inflow-outflow relations, the time to resolve a queue, as well as the associated waiting
times. The consistency with other and more complex traffic flow models is shown in
Ref.~\cite{Helbing2007}.

% ------------------------------------------------------------------------------------
% ------------------------------------------------------------------------------------
\subsection{Kirchhoff's law for the traffic dynamics at nodes}

Each node in a road network connects a number of incoming road sections denoted by the
index $i$ to a number of outgoing links denoted by $j$. Kirchhoff's law regarding the
conservation of flows at nodes requires that the flow arriving at an outgoing link $j$
equals the sum of the fractions $\alpha_{ij}(t)$ of the departure flows $Q_i^\dep(t)$
from the incoming links $i$, i.e.
\begin{equation}
    Q_j^\arr(t) = \sum\nolimits_i \alpha_{ij}(t) \, Q_i^\dep(t)
    \qquad \mbox{for all $j$ and $t$.}
\end{equation}
The turning fractions $\alpha_{ij}(t) \geq 0$ with $\sum_j \alpha_{ij} = 1$ are
normalized and may be time-dependent, as route choice and travel activities can change in
the course of day \cite{Janson1991,Helbing2001,Bowman2001,Lan2001}. By incorporating
limited arrival flows ($Q_j^\arr(t) \leq Q_j^\max$), it becomes obvious that a lack of
arrival capacity on a downstream link limits the departure flow on the upstream links,
which may eventually cause spill-back effects \cite{Daganzo1998}. A discussion of
concrete specifications of diverges and merges is provided in
Refs.~\cite{Helbing2005,Helbing2007}. For the dynamics of shock fronts propagating
through such network nodes, see Refs.~\cite{Herty2004,Garavello2006}.

When a traffic flow enters or crosses another one, i.e. at merging or intersection nodes,
the competing traffic flows tend to obstruct each other, which often leads to an
inefficient usage of intersection capacities \cite{Troutbeck1997,Helbing2006c}. Traffic
lights can serve to coordinate incompatible traffic flows and to increase the overall
performance. For traffic flows served by a green light, we assume in the following that
the outflow from a queue is only limited by the saturation flow $Q_i^\max$. That is,
throughout this paper, outflows will not be obstructed by other flows or by spill-backs
from downstream road sections.

A general approach to model the switching of traffic lights is to regulate the outflow of
an incoming road section $i$ with a ``permeability'' pre-factor $\gamma_i(t)$, which
alternates between 0 and 1 corresponding to a red and green traffic light, respectively
\cite{Helbing2007}. Three different regimes can be distinguished: (i) If the traffic
light is red, the outflow is zero. (ii) When the traffic light has switched to green, the
vehicle queue discharges at a more or less constant rate, the saturation flow $Q_i^\max$
\cite{Webster1958}. (iii) If the traffic light remains green after the queue has
dissolved, vehicles leave link $i$ at the same rate $Q_i^\exp(t) = Q_i^\arr(t-L_i/V_i)$
at which they enter it, delayed by the free travel time $L_i/V_i$. Together with
Eq.~(\ref{eq:n}), one obtains an ordinary differential equation for the temporal
evolution of the queue length $n_i(t)$:
\begin{equation}
    \label{eq:hybrid}
    \frac{dn_i}{dt} = \left\{ \begin{array}{ll}
        Q_i^\exp(t) &\mathrm{if\;} \gamma_i(t) = 0 \\
        Q_i^\exp(t) - Q_i^\max &\mathrm{if\;} \gamma_i(t) = 1 \mathrm{\;and\;}  n_i(t) > 0 \\
        0 &\mathrm{if\;} \gamma_i(t) = 1 \mathrm{\;and\;}  n_i(t) = 0.
    \end{array} \right.
\end{equation}
The above model allows us to characterize the queueing process at a signalized road
section as a nonlinear hybrid dynamical system \cite{Savkin2002}, i.e. a system of
equations containing both, continuous and discrete state variables. The transition from
regime (ii) to regime (iii), i.e. the transition from congested to free traffic is a
result of the particular arrival flow and cannot directly be controlled by the traffic
light. Thus, a complete formulation of the hybrid dynamical system requires to anticipate
the time point at which a queue will be cleared \cite{Lammer2007a}. This as well as the
switching losses due to reaction times and finite accelerations will be addressed in the
following section.

% ------------------------------------------------------------------------------------
% ------------------------------------------------------------------------------------
% ------------------------------------------------------------------------------------
\section{Anticipation of traffic flows and platoons}
\label{sec:Anticipation}

For a flexible traffic light control to be efficient, it is essential to {\it anticipate
the vehicle flows} as good as possible (see \ref{sec:limitedHorizon}). In
Ref.~\cite{Lammer2007a}, we have proposed a framework to predict the effects of starting,
continuing, or terminating service processes on future waiting times. The main results
are shortly summarized in the following and serve as the basis for deriving optimal
switching rules in Sec.~\ref{sec:Prio}.

\begin{figure}
    \includegraphics[width=\figwidth]{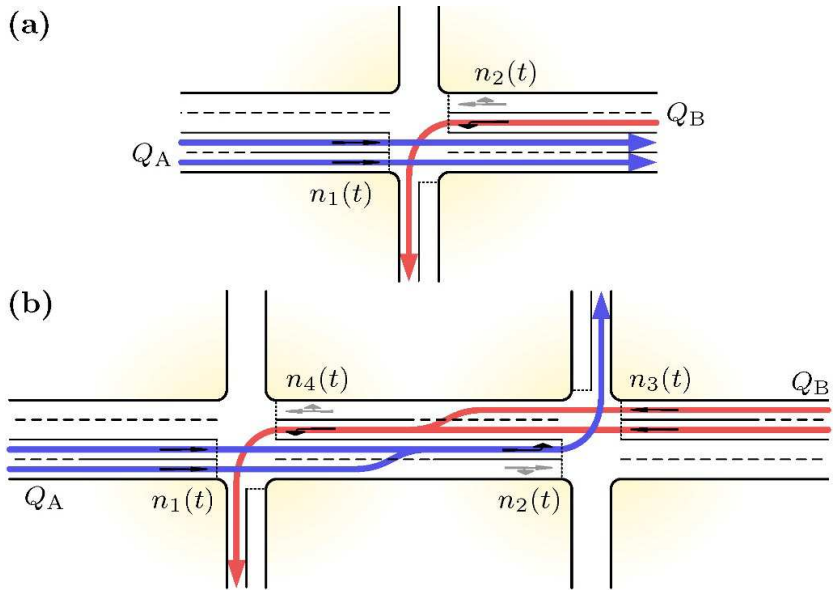}
    \caption{
(a) Isolated intersection with two incompatible traffic streams A and B. In this case, a
suitable clearing policy is both optimal and stable (see \ref{sec:dynamicInstabilities}).
(b) Combination of two intersections of the kind displayed in subfigure (a), forming a
non-acyclic road network (see \ref{sec:dynamicInstabilities}). It is interesting that,
even when each of the intersections behaves stable in isolation, the road network might
behave dynamically unstable under identical inflow conditions (see Fig.~\ref{fig:kumar}
and Ref. \cite{Kumar1990}).}
    \label{fig:intersections}
\end{figure}

\begin{figure}
    \includegraphics[width=\figwidth]{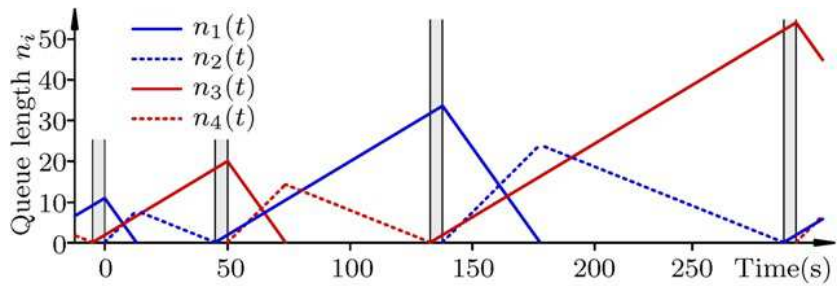}
    \caption{
Time-dependent queue lengths for the non-acyclic road network shown in
Fig.~\ref{fig:intersections}(b), assuming a clearing policy that behaves optimally at the
isolated intersection illustrated in Fig. ~\ref{fig:intersections}(a). The queue lengths
diverge due to dynamic instability. (For an explanation of the clearing policy see
\ref{sec:dynamicInstabilities}.) The reason for this instability lies in the inefficient
usage of service capacities during the time periods from 20 to 45 seconds, from 70 to 130
seconds, and so on. During this time, the traffic lights extend the green time for
streets 1 and 3 where the vehicle queues have already been cleared, while the other
streets are ``being starved of input'', using the words of Kumar and Seidman
\cite{Kumar1990}. }
    \label{fig:kumar}
\end{figure}

Note, however, that there are fundamental limits to the prediction of traffic flows (see
\ref{sec:Coordination}): Already very small networks with very simple switching rules can
produce a complex and potentially chaotic traffic dynamics (see \ref{sec:chaos}).
Moreover, coordination problems between traffic flows and their service may cause an
inefficient usage of intersection capacities and, thereby, spill-back effects and related
dynamic instabilities (see Figs.~\ref{fig:intersections}, \ref{fig:kumar} and
\ref{sec:dynamicInstabilities}). These can sometimes be quite unexpected and imply that
plausible optimization attempts may fail due to non-linear feedback effects. Details are
discussed in the Appendix.

% ------------------------------------------------------------------------------------
% ------------------------------------------------------------------------------------
\subsection{Service process and setup times}

The safe operation of traffic lights requires that, before switching to green for the
traffic flow of $i$, all other incompatible traffic flows have been stopped and all
corresponding vehicles have already left the conflict area. This will be considered in
our model by introducing setup times: If some traffic flow $i$ is selected for service,
its traffic light does not switch to green before the corresponding setup (or intergreen)
time $\tau_i^0$ has elapsed \cite{Smith2001}. The value of $\tau_i^0$ has to be chosen
according to safety considerations and usually lies in the range between 3 to 8 seconds.
Please note that $\tau_i^0$ also includes the amber time period, which takes into account
reaction delays and delays by finite acceleration. Therefore, the setup time $\tau_i^0$
reflects all time losses associated with the start of service for vehicles on link $i$.
As depicted in Fig.~\ref{fig:anticipation}(c), a service process can be divided into
three successive states: the setup, the clearing of the queue, and the green time
extension. The traffic light is green only in the latter two states.

% ------------------------------------------------------------------------------------
% ------------------------------------------------------------------------------------
\subsection{Green time required to clear a queue}

\begin{figure}
    \includegraphics[width=\figwidth]{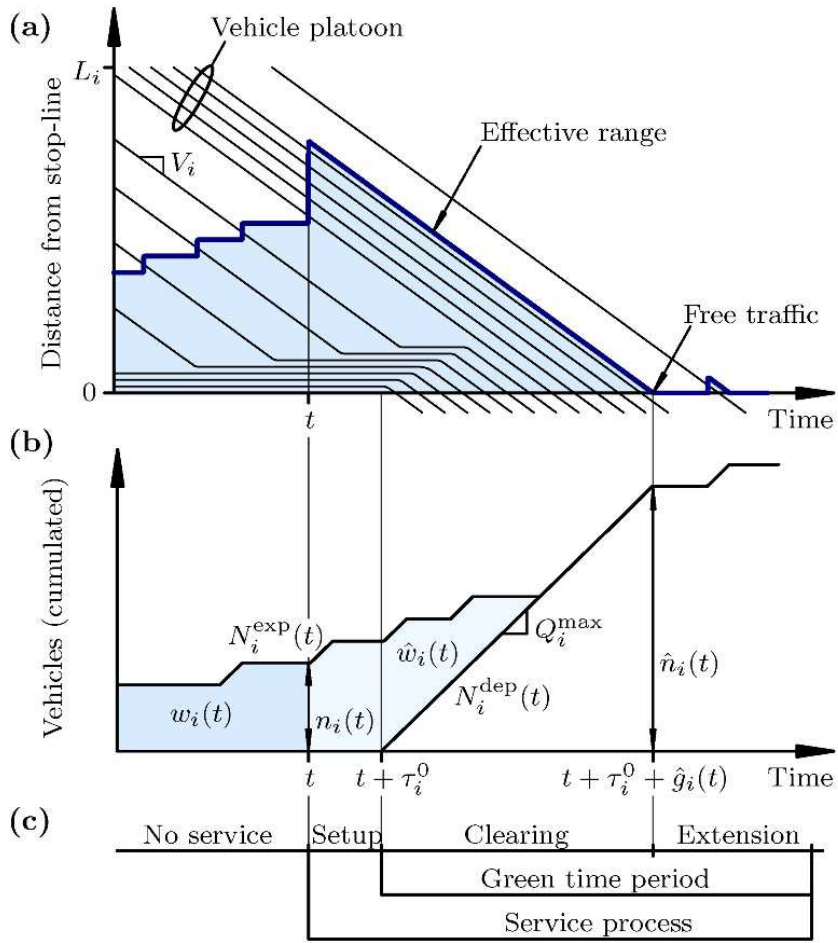}
    \caption{
(a) Trajectories and (b) cumulated number of vehicles on a road section $i$, and (c)
different states of the service process. The service process starts early enough to serve
a platoon of five vehicles in a green-wave manner, i.e. without stopping the vehicles.
The precise timing results from a short-term anticipation \cite{Lammer2007a} based on the
time series $N_i^\exp(t)$ and $N_i^\dep(t)$ (i.e. the cumulated number of vehicles that
could have reached the stop line in free traffic as compared to the number that has
actually have left the road section, see Eqs.~(\ref{eq:Nexp}) and (\ref{eq:Ndep})).
Whereas the current waiting time $w_i(t)$ grows with the number of vehicles $n_i(t)$
being delayed (see Eq.~(\ref{eq:n})), the expected future waiting time $\hat w_i(t)$
grows with the expected number of vehicles $\hat n_i(t)$ to be served in the subsequent
``clearing'' state (see Eqs.~(\ref{eq:dWdT}) and (\ref{eq:DeltaW})). The value of $\hat
n_i(t)$ as well as the required green time $\hat g_i(t)$ for clearing the queue are
determined by Eqs.~(\ref{eq:hatg}) and (\ref{eq:hatn}). A platoon is served in a
green-wave manner, if the start of the service process is initiated by the
platoon-related jump in $\hat n_i(t)$, or what is more illustrative, by the sudden
increase of the effective range (see text).}
    \label{fig:anticipation}
\end{figure}

For the flexible control of traffic lights it is of fundamental importance to anticipate
the amount of green time $\hat g_i(t)$ required for clearing the queue in road section
$i$, given the service starts or is being continued at the current time point $t$.
Obviously, $\hat g_i(t)$ does not only depend on the current queue length $n_i(t)$, but
also on the number of vehicles joining the queue during the remaining setup time
$\tau_i(t)$, and while the queue is being cleared. The queue of delayed vehicles has
fully dissolved at the time point $t + \tau_i(t) + \hat g_i(t)$, which is defined by the
requirement that the number of vehicles having left the road section by that time is
equal to the number of vehicles that have reached the stop-line. This corresponds to the
left- and right-hand side, respectively, of the following equation:
\begin{equation}
    \label{eq:hatg}
    N_i^\dep(t) + \hat g_i(t) \, Q_i^\max = N_i^\exp\big(t + \tau_i(t) + \hat g_i(t) \big).
\end{equation}
The value of $\hat g_i(t)$ shall be the largest possible solution of Eq.~(\ref{eq:hatg}),
which can be easily obtained with standard bisection methods \cite{Arfken1995}. The
second term in Eq.~(\ref{eq:hatg}) represents the number of vehicles that are expected to
leave the road section at the saturation flow rate $Q_i^\max$, and shall be denoted by
$\hat n_i(t)$, i.e.
\begin{equation}
    \label{eq:hatn}
    \hat n_i(t) = \hat g_i(t) \, Q_i^\max.
\end{equation}
A detailed derivation and discussion of the dynamics of $\hat g_i(t)$ and $\hat n_i(t)$
in different dynamical regimes is provided in Ref.~\cite{Lammer2007a}. $\hat n_i(t)$
captures all those vehicles
\begin{itemize}
    \item already waiting in the queue,
    \item joining the queue during setup or clearing, and
    \item arriving as a platoon immediately after the queue is cleared.
\end{itemize}
It particularly considers jumps to a higher value, when a platoon could be served in a
green-wave manner, i.e. without stopping. The magnitude of the jump is equal to the size
of the platoon. Before the platoon arrives at the stop-line, the formula reserves exactly
as much time as needed to perform the setup and to clear the queue of waiting vehicle.
Thus, the above anticipation model provides us with a mechanism that establishes green
waves. In order to visualize the underlying principle, Fig.~\ref{fig:anticipation}(a)
plots the so-called effective anticipation range, which includes the $\hat n_i(t)$
vehicles.

Note that, when the effective range extends $(\tau_i(t) + \hat g_i(t))V_i$ meters from
the stop-line, all vehicles within that range will reach the stop-line \emph{before} the
queue is being cleared at time point $t + \tau_i(t) + \hat g_i(t)$. These vehicles will,
thus, be served within the ``clearing'' state of a \emph{subsequent} service process.

% ------------------------------------------------------------------------------------
% ------------------------------------------------------------------------------------
\subsection{Waiting time anticipation}

Obviously, we would like to be able to decide whether to continue a service process or
start another one is more profitable in terms of saving waiting time. Therefore, the
above anticipation concept shall now be used to forecast the total waiting time $\hat
w_i(t)$ of all vehicles on road section $i$ up to the end of the subsequent ``clearing''
state (see Fig.~\ref{fig:anticipation}(b)). According to Ref.~\cite{Lammer2007a}, we have
\begin{equation}
    \label{eq:dWdT}
    \frac{d \hat w_i}{dt} =
    \left\{ \begin{array}{ll}
        \hat n_i(t) & \mbox{if $i$ is not served}\\
        0 & \mbox{during the entire service process.}
    \end{array} \right.
\end{equation}
That is, any delay $dt$ in the start of service will cause an additional delay $dt$ for
each of the expected vehicles. Interestingly, $\hat w_i(t)$ does not change anymore
during the service process, because the corresponding value has already been anticipated
before. However, it will change again as soon as the service process is terminated. At
the same point in time, the anticipated waiting time $\hat w_i(t)$ will also increase by
the additional amount $\Delta \hat w_i(t)$ due to the fact that the next green time
cannot start before performing a new setup, which takes a time period $\tau_i^0$. This
additional, setup-waiting time is given by
\begin{equation}
    \label{eq:DeltaW}
    \Delta \hat w_i(t) = Q_i^\max \int_{\tau_i(t)}^{\tau_i^0} \hat g_i(t, \tau') d \tau' \,,
\end{equation}
where $\hat g_i(t, \tau)$ corresponds to the solution of Eq.~(\ref{eq:hatg}), given a
remaining setup time of $\tau'$. The above Eqs.~(\ref{eq:dWdT}) and~(\ref{eq:DeltaW})
allow one to anticipate the costs of delaying or terminating a service process in terms
of expected future waiting times. To underline the particular importance of this result,
we would like to point out the direct relation between $n_i(t)$ and $\hat n_i(t)$: While
$n_i(t)$ is the growth rate of the \emph{current} waiting time $w_i(t)$ according to
Eq.~(\ref{eq:n}), $\hat n_i(t)$ is the growth rate of the \emph{expected} future waiting
time $\hat w_i(t)$ for a traffic flow $i$ that is not being served. This fundamental
similarity allows us to easily transfer conventional control schemes, which have
originally been developed to operate on $n_i(t)$, to the variables of our anticipation
model.

% ------------------------------------------------------------------------------------
% ------------------------------------------------------------------------------------
% ------------------------------------------------------------------------------------
\section{Conventional and self-organized traffic light control}
\label{sec:TrafficLightControl}

% ------------------------------------------------------------------------------------
% ------------------------------------------------------------------------------------
\subsection{The classical control approach and its limitations}
\label{sec:Review}

The optimal control of switched network flows is known to be an NP-hard problem
\cite{Papadimitriou1999}, which means that the time required to find an optimal solution
grows faster than polynomially with the network size (number of nodes). This NP-hardness
has two major implications: First, traffic light controls for road networks are usually
optimized off-line for certain standard situations (such as the morning or afternoon rush
hours, sports events, evening traffic, weekends, etc.), and applied under the
corresponding traffic conditions. Second, todays control approaches are predominantly
centralized and based on the application of pre-calculated periodic schedules, some
parameters of which may be adaptively adjusted (for a discussion of the related traffic
engineering literature see Ref.~\cite{Porche1996,Papageorgiou2003}). That is,
coordination is reached by applying a common cycle time to all intersections or multiples
of a basic frequency \cite{Webster1958}. This frequency is normally set by the most
serious bottleneck. For capacity reasons (to minimize inefficiencies due to switching
times), the frequency is reduced at high traffic volumes, but it is limited by a maximum
admissible cycle time. Apart from the cycle time, the order and relative duration of
green phases (the ``split''), and the time shifts between neighboring traffic lights
(``offsets'') are optimized for assumed boundary conditions (in- and outflows). The
resulting program usually serves each traffic flow once during the cycle time, and it is
repeated periodically. So-called ``green waves'' are implemented by suitable adjustment
of green phases and time shifts. They usually prioritize a unidirectional main flow (e.g.
in- or out-bound rush-hour traffic in ``arterials'') \cite{McDonald1991}.

Some obvious disadvantages of this classical control approach are:
\begin{enumerate}
\item
In order to cope with variations of the inflow, green times are often longer than needed
to serve the average number of arriving vehicles (otherwise excessive waiting times may
occur due to multiple stops in front of the same red light). This causes unnecessarily
long waiting times for incompatible flow directions.
\item
At intersections with small utilization, the cycle time is typically much longer than
required (or the cycle is uncoordinated with the intersection constituting the major
bottleneck). Moreover, traffic lights tend to cause avoidable delays during times of
light traffic (e.g. at night).
\item
A coordination through ``green waves'' is applicable to one traffic corridor and flow
direction only, while they tend to obstruct opposite, crossing, and merging flows.
\item
Due to the considerable variation of traffic flows and turning fractions from one minute
to another, the traffic light schedule is optimized for an \emph{average} situation which
is never met exactly, while it is not optimal for the \emph{actual} traffic situation.
\end{enumerate}

% ------------------------------------------------------------------------------------
% ------------------------------------------------------------------------------------
\subsection{Real-time heuristics based on a self-organized prioritization strategy}
\label{sec:Heuristics}

To overcome the before mentioned disadvantages, we propose to perform a heuristic on-line
optimization that flexibly adapts to the {\it actual} traffic situation at \emph{each}
time and place. If this heuristic reaches, on average, say 95\% of the performance of the
theoretically optimal solution, it is expected to be superior to the pre-determined,
100\% best solution for an {\it average} traffic situation that never occurs exactly.
Moreover, finding the one, 100\% best traffic light control for a given, time-dependent
situation is numerically so demanding that it requires off-line optimization, while
solutions reaching, say, 95\% of the optimal performance can be determined in real time.
As there are typically {\it several} alternative solutions of high, but not optimal
performance, it is also possible to select a solution that is particularly well adjusted
to the {\it local} traffic conditions.

In the following, we will specify a heuristics for a decentralized, real-time traffic
light control. In order to reach a superior performance as compared to a simple, cyclical
fixed-time control (see Sec.~\ref{sec:Simulation}), our self-organized prioritization
approach combines an optimizing strategy (see Sec.~\ref{sec:Prio}) with a stabilizing one
(see Sec.~\ref{sec:Stab}).

\begin{figure}
    \includegraphics[width=\figwidth]{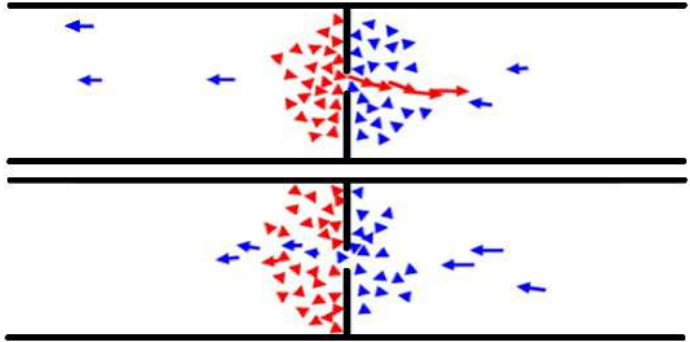}
    \caption{
Pedestrians flows at a narrow bottleneck behave almost as if they were controlled by
traffic lights (after Helbing and Moln\'{a}r \cite{Helbing1995}). }
    \label{fig:pedestrians}
\end{figure}

Our control concept is inspired by the observation that pedestrian counter-flows at
bottlenecks show a self-organized oscillation of their passing direction (see
Fig.~\ref{fig:pedestrians}), as if the pedestrians were controlled by traffic lights
\cite{Helbing1995,Helbing2001a,Helbing2006c}. In pedestrian flows, the self-organized
oscillations result from pressure differences between the waiting crowds on both sides of
the bottleneck. Pressure builds up on the side where more and more pedestrians have to
wait, while it is reduced on the side where pedestrians manage to pass the bottleneck.
The passing direction changes, when the pressure on one side exceeds the pressure on the
other side by a sufficient amount.

Intersections may also be viewed as bottlenecks, but with more than two flows competing
for the available service capacity. Therefore, our idea is to transfer the above
described self-organizing principle to urban vehicular traffic.

% ------------------------------------------------------------------------------------
% ------------------------------------------------------------------------------------
\subsection{Optimization strategy}
\label{sec:Prio}

We define ``pressures'' by dynamic priority indices $\pi_i(t)$ such that the traffic
lights of an intersection give a green light to the traffic flow $i$ with highest
priority. For the mathematical formulation of the dynamic prioritization rule, let us
store the argument $i$ in a decision variable $\sigma(t)$ as follows:
\begin{equation}
    \label{eq:maxPi}
    \sigma(t) = \arg\;\max_i \; \pi_i(t).
\end{equation}
Priority-based scheduling has been studied in the context of queueing theory
\cite{Righter2002,Oyen1992,Rothkopf1984,Serres1991,Panwalkar1977,Kumar1994,Harrison1975,Balachandran1970}.
It has been stated, that ``there are no undiscovered priority index sequencing rules for
minimizing total delay costs'' \cite{Rothkopf1984}. However, the considered
prioritization strategies were restricted to functions of the \emph{current} queue
length, i.e. to the number of vehicles that have \emph{already} been stopped
\cite{Oyen1992,Duenyas1996}. In contrast, our anticipation model (see
Sec.~\ref{sec:Anticipation}) allows one to predict {\em future} arrivals and to
generalize these strategies to serving platoons without any previous stops, i.e. in a
``green wave'' manner. For simplicity, we will assume in the following, that route-choice
is non-adaptive (i.e. the turning fractions $\alpha_{ij}(t)$ are known) and also that all
traffic flows at the intersections are conflicting (i.e. only one traffic flow can be
served at a time).

Our goal is to derive a formula for the priority index $\pi_i$ such that switching rule
(\ref{eq:maxPi}) minimizes the total waiting time. However, the optimization horizon is
limited to those vehicles, whose future waiting time directly depends on the
\emph{current} state of the traffic lights, i.e. the expected $\hat n_i$ vehicles
captured within the effective range (see Fig.~\ref{fig:anticipation}(a)). Later arriving
vehicles are neglected as long as they are beyond the anticipation horizon, but they are
taken into account by the dynamic re-optimization early enough to serve them by a green
wave if this is possible.

\begin{figure}
    \includegraphics[width=\figwidth]{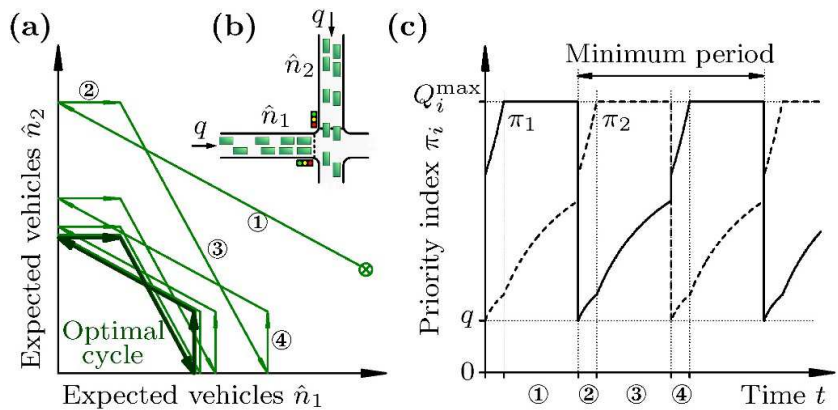}
    \caption{
(a) Convergence of the trajectories $(\hat n_1, \hat n_2)$ to the optimal limit cycle at
an intersection with two identical traffic flows with constant inflow rate $q$, see (b).
(c) Periodic time series of the priority indices $\pi_1$ and $\pi_2$ associated with the
optimal cycle. (\COne means clearing street 1, \CTwo setup for street 2, \CThree clearing
street 2, and \CFour setup for street 1)}
    \label{fig:prio}
\end{figure}

In case of no further arrivals, Rothkopf and Smith \cite{Rothkopf1984} showed that the
optimal order of serving traffic flows is unique and can be determined by comparing
priorities among \emph{pairs} of competing traffic flows. This allows us to derive the
optimal specification of the priority index $\pi_i$ by studying an intersection of only
\emph{two} competing traffic flows 1 and 2, as depicted in Fig.~\ref{fig:prio}(b). For
the current time point $t$, we assume the remaining setup times $\tau_1$ and $\tau_2$,
the anticipated number of vehicles $\hat n_1$ and $\hat n_2$, and the required green
times $\hat g_1$ and $\hat g_2$ to be given. We assume that, initially, traffic flow 1 is
being selected for service, i.e. $\sigma = 1$. In this scenario, the controller has two
options:
\begin{enumerate}[1.]
\item
to finish serving flow 1 before switching to flow 2 or
\item
to switch to flow 2 immediately, at the cost of an extra setup for switching back to flow
1 later on.
\end{enumerate}
The optimal control decision is derived by calculating the total increase in the
anticipated waiting time for each option. Following the first option requires to continue
serving flow 1 for $\tau_1 + \hat g_1$ seconds. According to Eq.~(\ref{eq:dWdT}), the
anticipated waiting time of traffic flow 2 grows at the rate $\hat n_2$, while it remains
constant for the traffic flow 1 under service. Since it does also not change after queue
1 has been cleared and while flow 2 is being served, the total increase of the
anticipated waiting time associated with the first option would be
\begin{equation}
    \label{eq:optionA}
    (\tau_1 + \hat g_1) \, \hat n_2 \,.
\end{equation}

When selecting the second option, according to Eq.~(\ref{eq:DeltaW}) the termination of
the service of traffic flow 1 causes the anticipated waiting time to increase by the
amount $\Delta \hat w_1$, which reflects the extra waiting time associated with the setup
for switching back later. While serving traffic flow 2 for $\tau_2 + \hat g_2$ seconds,
the anticipated waiting time grows further at the rate $\hat n_1$. Altogether, its total
increase would be
\begin{equation}
    \label{eq:optionB}
    \Delta \hat w_1 + (\tau_2 + \hat g_2) \, \hat n_1 \,.
\end{equation}
Thus, it is optimal to continue serving traffic flow 1 as compared to switching to flow 2
if
\begin{equation}
    \label{eq:OptimalCondition}
    (\tau_1 + \hat g_1) \, \hat n_2
    \;<\;
    \Delta \hat w_1 + (\tau_2 + \hat g_2) \, \hat n_1
    \,.
\end{equation}
The above optimality criterion allows us to define priority indices $\pi_1$ and $\pi_2$
by separating the corresponding variables. For this, we rewrite
Eq.~(\ref{eq:OptimalCondition}) in the following way
\begin{equation}
    \label{eq:pi1pi2}
    \pi_1 := \frac{\hat n_1}{\tau_1 + \hat g_1}
    \;>\;
    \frac{\hat n_2}{\Delta \hat w_1 / \hat n_1 + \tau_2 + \hat g_2} =: \pi_2
    \,.
\end{equation}
Each side of this inequality defines a priority index $\pi_i$. With this definition, the
priority $\pi_1$ for traffic flow 1 is a function of its own variables only.
Interestingly, $\pi_2$ has the same dependence on its own variables, but it additionally
depends on the term $\Delta \hat w_1 / \hat n_1$. Before we can derive a general formula
for the priority $\pi_i$ of \emph{any} traffic flow $i$, we must first clarify the role
of this extra term. In general, the expression $\Delta \hat w_\sigma / \hat n_\sigma$
reflects the penalty for terminating the current service process, where $\sigma$ stands
for the traffic flow being served. As it follows from Eq.~(\ref{eq:DeltaW}), the value of
$\Delta \hat w_\sigma / \hat n_\sigma$ ranges from $0$ to $\tau_\sigma^0$ and thus
represents the additional waiting time $\Delta \hat w_\sigma$ due to the extra setup for
switching back, averaged over all corresponding vehicles $\hat n_\sigma$. Since the
penalty for switching from $\sigma$ to $i$ applies only to those traffic flows $i \neq
\sigma$ not being served, we can introduce the general penalty term $\tau_{i,
\sigma}^\pen$ as follows:
\begin{equation}
    \label{eq:penalty}
    \tau_{i, \sigma}^\pen = \left\{
        \begin{array}{ll}
            \Delta \hat w_\sigma / \hat n_\sigma &\mathrm{\;if\;} i \neq \sigma \\
            0 &\mathrm{\;if\;} i = \sigma \,.
        \end{array}
    \right.
\end{equation}
With this notation we can introduce the general definition of the priority index $\pi_i$
as
\begin{equation}
    \label{eq:pi}
    \pi_i =
    \; \frac{\hat n_i}{\tau_{i, \sigma}^\pen + \tau_i + \hat g_i}
    \,.
\end{equation}
This is fully compatible with the optimality criterion~(\ref{eq:pi1pi2}). To interpret
the result, the priority index $\pi_i$ relates to the anticipated average service rate,
i.e. the anticipated number $\hat n_i$ of vehicles expected to be served during the time
period $\tau_i + \hat g_i$. In contrast to conventional priority specifications derived
from the so-called $\mu c$ rule \cite{Oyen1992,Serres1991,Duenyas1996,Baras1985},
specification (\ref{eq:pi}) is novel in two fundamental aspects: First, its dependence on
the predicted variables $\hat n_i$ and $\hat g_i$ allows one to anticipate future
arrivals (see Sec.~\ref{sec:Anticipation}). Second, it takes into account both first- and
second-order switching losses, i.e. the setup times for switching to another traffic flow
as well as for switching back, represented by $\tau_i$ and $\tau_{i, \sigma}^\pen$,
respectively.

Instead of clearing existing queues in the most efficient way, our anticipative
prioritization strategy aims at minimizing waiting times. This prevents queues to form
and causes green waves to emerge automatically, whenever this saves overall waiting time
at the intersection. The underlying mechanism relates to the fact that the values of
$\hat n_i$ and $\hat g_i$ jump to a higher value as soon as the first vehicle of a
platoon enters the dynamic anticipation horizon (see Sec.~\ref{sec:Anticipation}).

Whether a platoon is being served by a green wave or not finally depends, of course, on
the overall traffic situation at the local intersection. While our previous
considerations applied to vehicle queues of given length, the same prioritization rule
shows a fast, exponential convergence to the optimal traffic light cycle also for
continuous inflows (see Fig.~\ref{fig:prio}). However, a local optimization of each
single intersection must not necessarily imply global optimality for the entire network
\cite{Lefeber2006,Gartner1975,Gartner1975a,Gershenson2007}, as dynamic instabilities
cannot be excluded (see \ref{sec:dynamicInstabilities}). Thus, our self-organized traffic
light control must be extended by a stabilization strategy.

% ------------------------------------------------------------------------------------
% ------------------------------------------------------------------------------------
\subsection{Stabilization strategy}
\label{sec:Stab}

We call a traffic light control ``stable'', if the queue lengths will always stay
finite~\cite{Perkins1989}. Of course, stability requires that the traffic demand does not
exceed the intersection capacities. Nevertheless, the short-sightedness of locally
optimizing strategies could lead to an inefficient use of capacity, e.g. because of too
frequent switching or too long green time extensions. This problem can be illustrated
even by analytical examples, see Refs.~\cite{Kumar1990,Duenyas1996,Burgess1997}. For a
discussion see Sec.~\ref{sec:Anticipation} and \ref{sec:dynamicInstabilities}. As a
consequence, even when the traffic demand is far from being critical, there is a risk
that vehicle queues grow longer and longer and eventually block traffic flows at upstream
intersections \cite{Daganzo2007}.

In order to stabilize a switched flow network, one may implement local supervisory
mechanisms \cite{Kumar1990}. The function of such mechanisms is to observe the current
traffic condition and to assign sufficiently long green times before queues become too
long. Maintaining stability is more of a resource allocation (green time assignment)
rather than a scheduling problem.

Our proposal is to complement the prioritization rule (\ref{eq:pi}) by the following
stabilization rule: We define an ordered priority set $\Omega$ containing the arguments
$i$ of all \emph{those} traffic flows, that have been selected by the supervisory
mechanism and, thus, need to be served soon in order to maintain stability. Furthermore,
the argument $i$ of a crowded link $i$ \emph{joins} the set $\Omega$ as soon as more than
some critical number $n_i^\crit$ of vehicles is waiting to be served. It is
\emph{removed} from the set after the queue was cleared, i.e. $n_i = 0$, or after a
maximum allowed green time $g_i^\max$ was reached. Elements included in the set $\Omega$
are served on a first-come-first-serve basis. As long as $\Omega$ is not empty, the
control strategy is to always serve the traffic flow corresponding to the first element
(head) of $\Omega$. If $\Omega$ is empty, the traffic lights follow the prioritization
rule (\ref{eq:maxPi}).

% ------------------------------------------------------------------------------------
% ------------------------------------------------------------------------------------
\subsection{Combined strategy}
\label{sec:Combined}

Our new control strategy can be summarized as follows:
\begin{equation}
    \label{eq:Combined}
    \sigma = \left\{ \begin{array}{ll}
        \mathrm{head} \; \Omega \quad& \mathrm{if}\; \Omega \neq \emptyset \\
        \arg \max_i \; \pi_i   \quad& \mathrm{otherwise}.
    \end{array}\right.
\end{equation}
It is, therefore, a combination of two complementary control regimes. Whereas the
optimizing regime (while $\Omega = \emptyset$) aims for minimizing waiting times by
serving the incoming traffic as quickly as possible, the stabilizing regime (while
$\Omega \neq \emptyset$) intervenes only if the optimizing regime fails to keep the queue
lengths below a certain threshold $n_i^\crit$. This means that, as long as the optimizing
regime itself exhibits the desired behaviour, i.e. as long as it is stable, the
stabilizing regime will \emph{never} intervene. If it needs to be activated for
particular traffic flows $i$ with $\hat n_i > n_i^\crit$, however, the control is handed
back to the optimizing regime as soon as the critical queues have been cleared.

Originally, such stabilizing supervisory mechanisms have been proposed for the control of
production and communication systems, e.g. in
Refs.~\cite{Kumar1990,Duenyas1996,Chang1992,Chang1994,Ram1993}. As such rules would,
however, not explicitly pay attention to the duration of red traffic lights, they would
not be suited for the application to urban road networks: Too long red times would
increase the risk of red-light violations and therefore also the risk of traffic
accidents \cite{Retting1997,Porter2000,Datta2000}. Thus, it is essential to have a good
model for the service intervals.

% ------------------------------------------------------------------------------------
\subsubsection{Service intervals}

In the following, we will specify the critical thresholds $n_i^\crit$ and the maximum
green times $g_i^\max$ such that the stabilization rule alone ($\sigma = \mathrm{head\,}
\Omega$) fulfills the following two safety requirements: Each traffic flow shall be
served
\begin{itemize}
\item[(S1)] once, on average, within a desired service interval $T > 0$ and
\item[(S2)] at least once within a maximum service interval $T^\max \geq T$.
\end{itemize}
These two parameters, $T$ and $T^\max$, are the only two adjustable parameters of our
control algorithm.

As service interval $z_i$, we define the time interval between two successive service
processes for the same traffic flow $i$. Accordingly, the service interval $z_i$ is the
sum
\begin{equation}
    \label{eq:z}
    z_i = r_i + \tau_i^0 + \hat g_i
\end{equation}
of the preceding red time of $r_i$, the setup time $\tau_i^0$, and the green time $\hat
g_i$ anticipated before the start of the service process. Thus, we can anticipate the
service interval $z_i$ before the corresponding service process starts. This allows us to
replace the critical threshold $n_i^\crit$ by a function $n_i^\crit(z_i)$ of the
anticipated service interval $z_i$.

\begin{figure}
    \includegraphics[width=\figwidth]{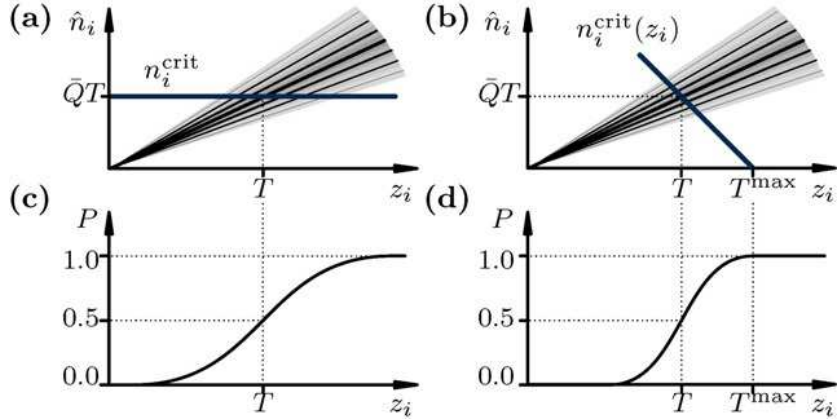}
    \caption{
Top: Anticipated number of vehicles $\hat n_i$ to be served within a service interval
$z_i$, following from the stochastic arrivals of the vehicles (fan of diagonal lines). If
the service process is started as soon as $\hat n_i$ exceeds the threshold function
$n_i^\crit(z_i)$ (thick line), the corresponding service interval is $z_i$.
$n_i^\crit(z_i)$ was specified as in Eq.~(\ref{eq:nCrit}) and plotted for $T^\max
\rightarrow \infty$ (left) and $T^\max < \infty$ (right). Bottom: Probability
distribution $P(Z \leq z_i)$. Because the probability for $z_i < T$ is 50\%, and for $z_i
< T^\max$ it is 100\%, our controller fulfills both safety requirements, (S1) and (S2).}
    \label{fig:erlang}
\end{figure}

Let us now study the statistical distribution of the service interval $z_i$ for a traffic
flow $i$ with random arrivals. Under the assumption that $n_i^\crit(z_i)$ is
non-increasing and the traffic flow is being served as soon as $\hat n_i \geq
n_i^\crit(z_i)$, we can make the following general statement: The probability $P(Z \leq
z_i)$ that the service interval $Z$ is shorter than $z_i$ is equal to the probability
that more than $n_i^\crit(z_i)$ vehicles arrive within a time interval $z_i$. The
probability distribution $P(Z \leq z_i)$ can be derived from a given function
$n_i^\crit(z_i)$ and a given stochastic model of the arrival process, for example using
the framework proposed in Refs.~\cite{Perry1999,Abdel-Hameed2000}.
Figure~\ref{fig:erlang} illustrates the distribution for two different threshold
functions $n_i^\crit(z_i)$.

From the above observations, we can now derive an appropriate specification of
$n_i^\crit(z_i)$. Most importantly, safety requirement (S1) can be fulfilled
independently of the particular arrival process. Following the above arguments, this mean
that $P(Z \leq T^\max)=1$ can be enforced by requiring
\begin{equation}
    \label{eq:maxServiceInterval}
    n_i^\crit(z_i) \leq 0
    \quad\mathrm{for\;}z \geq T^\max.
\end{equation}
Thus, no matter how few vehicles actually arrived, the corresponding traffic flow will be
served once within $T^\max$. One possible specification is
\begin{equation}
    \label{eq:nCrit}
    n_i^\crit(z_i) = \bar Q_i T \, \frac{T^\max - z_i}{T^\max - T},
\end{equation}
where $\bar Q_i$ denotes the average arrival rate. This specification satisfies condition
(\ref{eq:maxServiceInterval}), but also fulfills the safety requirement (S2). Within the
desired service interval $T$, there will, on average, arrive a number of $\bar Q_i T$
vehicles. This number, however, is equal to the critical threshold $n_i^\crit(z_i)$ for
an anticipated service interval of $z_i = T$. Thus, a service process is started
immediately when there are as many vehicles to serve, as there arrive on average within
the desired service time period $T$. Figure~\ref{fig:erlang} plots the distribution of
service intervals $z_i$ for different parameters of the threshold function
$n_i^\crit(z_i)$ according to specification (\ref{eq:nCrit}). Altogether, the probability
of having $z_i < T$ is 50\%, and the probability for $z_i < T^\max$ is 100\%.

Let us briefly discuss two limiting cases: (i) If $T^\max \rightarrow \infty$, the
threshold function $n_i^\crit(z_i) = \bar Q_i T$ becomes a horizontal line as depicted in
Fig.~\ref{fig:erlang}(a). This parameter choice corresponds to a fully vehicle-responsive
operation, where one does not care about the duration of the actual service interval.
(ii) If $T^\max \rightarrow T$, the threshold function becomes a vertical line at $z_i =
T$. This case, in contrast, corresponds to a pure fixed-time operation with cycle time
$T$, where the actual traffic situation is completely ignored. In between these two
limiting cases, i.e. for $T < T^\max < \infty$, the switching behaviour is both,
time-dependent and vehicle-responsive.

% ------------------------------------------------------------------------------------
\subsubsection{Sufficient stability condition}

To make our control concept complete, the last step is to specify the maximum allowed
green time $g_i^\max$ for traffic flow $i$ in the stabilization strategy. Once $g_i^\max$
is exceeded, the element $i$ is removed from $\Omega$, even if its queue has not been
fully cleared in this time. Obviously, $g_i^\max$ must be chosen large enough in order to
maintain stability \cite{Kumar1990}. In particular, serving an average number of $\bar
Q_i T$ vehicles requires to provide a green time of at least $T \bar Q_i / Q_i^\max$
seconds. On the other hand, serving all traffic flows one after the other for $\tau_i^0 +
g_i^\max$ seconds each should not take more than $T$ seconds in total. Therefore,
$g_i^\max$ must meet the constraints
\begin{equation}
    \label{eq:gMaxLowerLimit}
    g_i^\max \geq T \, \bar Q_i / Q_i^\max \quad \mathrm{for\;all\;} i
\end{equation}
and
\begin{equation}
    \label{eq:gMaxUpperLimit}
    \sum\nolimits_i \left( \tau_i^0 + g_i^\max \right) \leq T .
\end{equation}
In order to obtain a sufficient condition for the existence of stable solutions, one can
insert $g_i^\max$ from Eq.~(\ref{eq:gMaxLowerLimit}) into Eq.~(\ref{eq:gMaxUpperLimit}),
which leads to
\begin{equation}
    \label{eq:stabSufficient_zwischenschritt}
    \sum\nolimits_i \tau_i^0
    \leq
    \left( 1 - \sum\nolimits_i \bar Q_i / Q_i^\max \right) T \,.
\end{equation}
That is, the sum of setup times must be smaller than the fraction of the service period
$T$ not needed to serve arriving vehicles. This condition is consistent with the
condition of Savkin \cite{Savkin1998,Savkin1998a} for a general switched server queueing
system to be controllable. Condition~(\ref{eq:stabSufficient_zwischenschritt}) also
indicates that there is a lower threshold for the desired service time period $T$:
\begin{equation}
    \label{eq:stabSufficient}
    T \, \geq \, \frac{\sum_i \tau_i^0}{1 - \sum_i \bar Q_i / Q_i^\max} \,.
\end{equation}
Interestingly, the same threshold has been shown to be the shortest possible cycle for a
stable \emph{periodic} switching
sequence~\cite{Webster1958,Savkin2002,Savkin2003,Lammer2006}. Therefore, we can conclude
that our self-organized, non-periodic traffic control defined by Eq.~(\ref{eq:Combined})
is stable whenever there exists a stable \emph{fixed-time} control with cycle time $T$.

\begin{figure}
    \includegraphics[width=\figwidth]{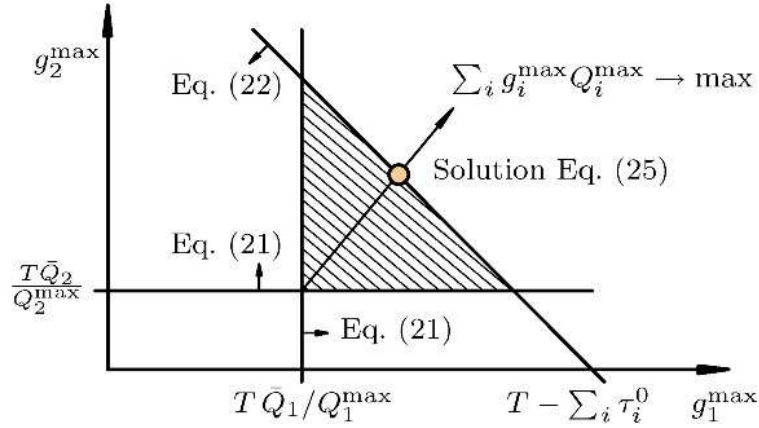}
    \caption{
Stable solutions for the maximum green times $g_i^\max$ lie within the simplex (shaded
area) constrained by Eqs.~(\ref{eq:gMaxLowerLimit}) and (\ref{eq:gMaxUpperLimit}). The
optimal values for $g_i^\max$ can be obtained from maximizing the throughput $\sum_i
g_i^\max Q_i^\max$. An easily computable explicit solution (circle) is given by
Eq.~(\ref{eq:gMax}).}
    \label{fig:lop}
\end{figure}

For a given desired service time period $T$ satisfying the stability
condition~(\ref{eq:stabSufficient}), the corresponding $g_i^\max$ values can be obtained
by solving an optimization problem. To minimize the average waiting times over an
interval $T$, one maximizes the overall throughput $\sum_i Q_i^\max g_i^\max$ as proposed
in Refs.~\cite{Gazis1964,Gazis2002,Schutter2002}. In order to solve this optimization
problem, however, it is necessary to know how much green time must be reserved for all
other traffic flows \cite{Lammer2007}. The determination of the exact optimum would
require to predict future arrivals over a prognosis horizon of about $T$ (i.e. normally
much longer than one minute). Because this is usually not possible (see
\ref{sec:limitedHorizon}), we suggest to determine a nearly optimal solution instead.
Setting
\begin{equation}
    \label{eq:gMax}
    g_i^\max = \frac{\bar Q_i}{Q_i^\max} \, T
    \, + \,
    \frac{Q_i^\max}{\sum_{i'} Q_{i'}^\max} \, T^\res
\end{equation}
(see the circle in Fig.~\ref{fig:lop}) satisfies both
constraints~(\ref{eq:gMaxLowerLimit}) and (\ref{eq:gMaxUpperLimit}). The first term on
the right-hand side of Eq.~(\ref{eq:gMax}) represents the minimum required green time $T
\bar Q_i / Q_i^\max$ according to Eq.~(\ref{eq:gMaxLowerLimit}). The second term adds a
fraction of the ``residual time'' $T^\res$ proportional to the corresponding saturation
flow $Q_i^\max$. Herein, the residual time is defined as
\begin{equation}
    \label{eq:TRes}
    T^\res = T \, \left( 1 - \sum\nolimits_i \bar Q_i / Q_i^\max \right) -
    \sum\nolimits_i \tau_{i}^0 \,,
\end{equation}
i.e. as the part of the service interval $T$ that is not necessarily needed for service
processes. (In other words, stability would still be guaranteed even if the traffic
lights would not serve any traffic flow for $T^\res$ seconds within the service interval
$T$. Thus, $T^\res$ relates to the free intersection capacity, which is here being used
to provide maximum possible green times if they are needed by the stabilization
strategy.)

% ------------------------------------------------------------------------------------
\subsubsection{Conclusion}

\begin{figure}
    \includegraphics[width=\figwidth]{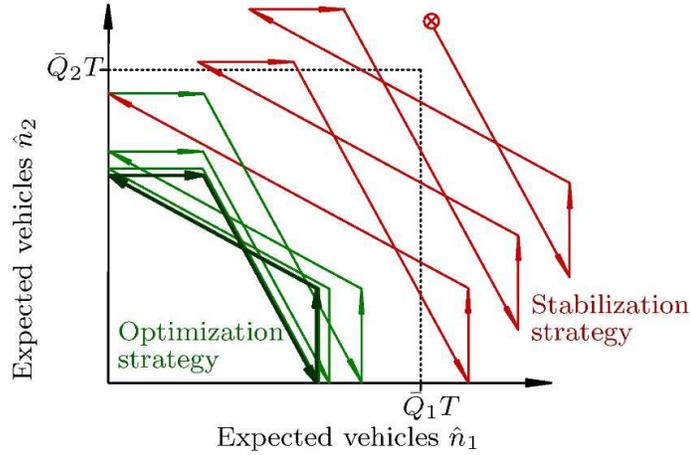}
    \caption{
Mutual, time-dependent interdependency of the expected number of vehicles $(\hat n_1,
\hat n_2)$ at an intersection with two identical traffic flows. The initial state
(crossed circle) corresponds to an over-saturated traffic condition. To clear the queues
as fast as possible, the stabilization strategy (red) minimizes switching losses by
serving each traffic flow exactly once within the desired service interval $T$. As soon
as the trajectories are below the critical threshold $n_i^\crit = \bar Q_i T$ defined by
Eq.~(\ref{eq:nCrit}) in the limit $T^\max \rightarrow \infty$, the optimization strategy
is being activated (green lines). The optimization strategy uses the available free
intersection capacity to converge towards the fastest possible switching sequence, which
is the optimum traffic light cycle in terms of travel time minimization.}
    \label{fig:stab}
\end{figure}

Our decentralized traffic light control strategy given by Eq.~(\ref{eq:Combined}) should
stabilize traffic flows in a road network as long as the traffic demands $\bar Q_i$ and
the desired service interval $T$ satisfy the sufficient stability condition
(\ref{eq:stabSufficient}). Interestingly, this condition is satisfied whenever there
exists a stable fixed-time control with cycle time $T$. Furthermore, both safety
requirements (S1) and (S2) are fulfilled under all circumstances, i.e. even for
over-saturated traffic conditions, where the eventual growth of vehicle queues is
unavoidable. In this case, the stabilization strategy serves the ingoing traffic flows
one after the other for $\tau_i^0 + g_i^\max$ seconds each. After the traffic situation
has relaxed, i.e. as soon as all queues can be cleared again within the desired service
interval $T$, the control is handed over to the optimization strategy. This uses the
available free intersection capacity $T^\res$ according to Eq.~(\ref{eq:TRes}) for
flexible switching sequences or green time extensions, i.e. for more frequent setups or
idling periods, as long as it helps to save waiting times. Such a scenario is illustrated
in Fig.~\ref{fig:stab}: At an initially over-saturated intersection, the stabilization
strategy manages to reduce the queue lengths, before it hands over to the optimization
strategy, which lets the queue lengths exponentially converge to the optimum cycle
associated with minimum waiting times.

% ------------------------------------------------------------------------------------
% ------------------------------------------------------------------------------------
% ------------------------------------------------------------------------------------
\section{Simulation of the self-organized traffic light control}
\label{sec:Simulation}

We have simulated the above control strategy (\ref{eq:Combined}) with the macroscopic
network flow model sketched in Sec.~\ref{sec:FlowModel}, using our short-term flow
anticipation algorithm (see Sec.~\ref{sec:Anticipation}). For comparison, the same has
been done with a car-following model within the microscopic simulation tool VISSIM
\cite{PTV}. This has resulted in qualitatively the same and quantitatively very similar
results, so that we do not show these duplicating results here.

For simplicity, our computer simulations assume that all traffic flows at the
intersections are incompatible, i.e. only one traffic flow can be served at a time. In
the following, we will report the corresponding simulation results and analyze the
performance of our control strategy.

% ------------------------------------------------------------------------------------
% ------------------------------------------------------------------------------------
\subsection{Operation modes at an isolated intersection}
\label{sec:SimOperationModes}

\begin{figure}
    \includegraphics[width=\figwidth]{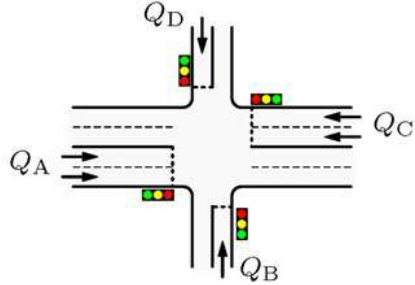}
    \caption{
Isolated intersection with four competing traffic flows.}
    \label{fig:singleIntersection}
\end{figure}

As a first test scenario, we study an isolated intersection with four traffic flows as
depicted in Fig.~\ref{fig:singleIntersection}. We are interested in the average total
queue length $\bar n = \langle \sum_i n_i \rangle$ in the steady state, i.e. over one
simulation hour. Whereas the inflow on the side streets was set to a constant volume of
$Q_\B = Q_\D = 180$ vehicles per hour, the inflow $Q_\A = Q_\C$ on the two-lane main
streets was varied. With a saturation flow rate of 1800 vehicles per hour and lane, we
had $Q_\A^\max = Q_\C^\max = 3600$ vehicles per hour and $Q_\B^\max = Q_\D^\max = 1800$
vehicles per hour. Furthermore, the setup times to switch between traffic flows were
$\tau_i^0 = 5$ seconds. With the control parameters $T=120$ seconds and $T^\max = 180$
seconds for the desired and the maximum service intervals, respectively, the sufficient
stability condition Eq.~(\ref{eq:stabSufficient}) was satisfied, if the utilization
\begin{equation}
    u = \sum\nolimits_i Q_i / Q_i^\max
\end{equation}
was less than 0.83. This means that our traffic light control was stable as long as the
average inflow on the main streets $Q_\A = Q_\C$ was less than 1140 vehicles per hour.

For different levels of saturation, our self-organized traffic light control exhibits
several distinct operation regimes:

% ------------------------------------------------------------------------------------
\subsubsection{Serving single vehicles at low utilizations}

\begin{figure}
    \includegraphics[width=\figwidth]{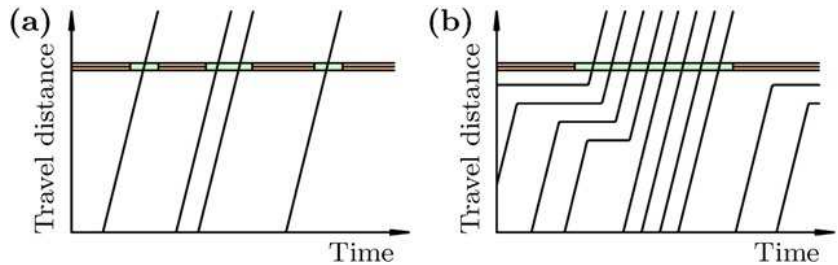}
    \caption{
Illustration of vehicle trajectories for different operation modes: (a) In the
low-utilization regime, the vehicles are served by a green light just upon their arrival
at the stop line (horizontal bar). Thereby, a stopping of the vehicles can be avoided.
(b) At higher utilizations, the formation of vehicle platoons is unavoidable. However,
serving vehicle platoons rather than maintaining the first-come-first-serve principle
allows one to minimize the average waiting time, as switching losses are reduced.}
    \label{fig:modes}
\end{figure}

In the low-utilization regime, traffic demand is considerably below capacity. A
minimization of the {\it average} waiting times is achieved by serving the vehicles just
upon their arrival, i.e. according to a first-come-first-serve principle. This operation
mode, which also minimizes {\it individual} travel times, is illustrated in
Fig.~\ref{fig:modes}(a).

% ------------------------------------------------------------------------------------
\subsubsection{Service of platoons at moderate utilizations}

As the traffic demand increases, several vehicles may arrive at the intersection at about
the same time, i.e. they may mutually obstruct each other. Some vehicles will have to
wait, which implies the formation of platoons. However, given a certain utilization
level, serving platoons becomes more efficient than applying the first-come-first-serve
principle (see Fig.~\ref{fig:modes}(b)): The reduction of switching losses by serving
platoons rather than single vehicles does not only reach a higher intersection capacity,
but also a minimization of the average travel times.

\subsubsection{Suppression of minor flows at medium utilizations}

In Fig.~\ref{fig:isoPeriod}, for utilizations $u$ between about 0.3 and 0.5 one can see
that the (multi-lane) main streets are served more frequently than the (one-lane) side
streets. That means, the interruption of the main flows by minor flows is suppressed,
which is again in favor of minimizing the average waiting times.

% ------------------------------------------------------------------------------------
\subsubsection{Flow stabilization at high utilizations}

\begin{figure}
    \includegraphics[width=\figwidth]{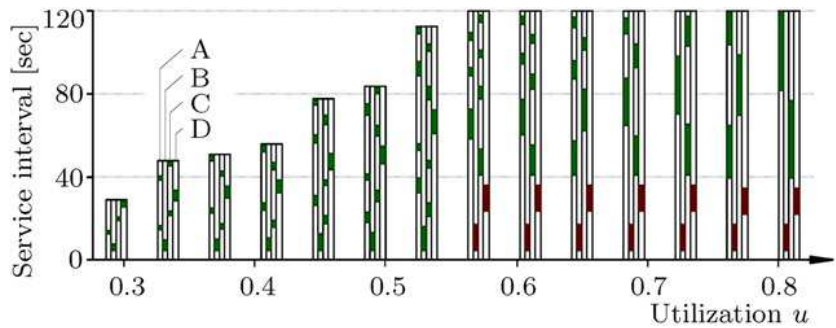}
    \caption{
Flexible switching sequences for different utilization levels $u$ over a complete service
interval of traffic flow B. The results were obtained by computer simulation of the
traffic flows at a single intersection after a transient time period. For details see
Sec. \ref{sec:SimIsolated}.}
    \label{fig:isoPeriod}
\end{figure}

For even higher utilizations $u$, our self-organized traffic light control does not
exclusively follow the travel time optimization strategy any longer: the side streets
would be served too rarely or too short. This becomes clear in Fig.~\ref{fig:isoPeriod}
for utilizations $u$ above 0.55. An efficient usage of the intersection capacity is now
reached by serving the side streets as soon as their vehicle queues have reached a
critical size. Thereby, the stabilization mechanism (see Sec.~\ref{sec:Stab}) ensures
that the safety-critical service interval of $T = 120$ seconds is never exceeded.
Interestingly, there have emerged switching sequences of higher periods, that is, it may
require several service intervals $T$ before a switching sequence repeats. Nevertheless,
all remaining capacity is still used to serve the main streets in the most flexible way,
i.e. by serving them as often as possible.

% ------------------------------------------------------------------------------------
%\subsubsection{Throughput maximization at saturated conditions}

%When the traffic demand exceeds capacity, which is the case for utilization $u > 0.83$,
%one faces the congested queue-dominated regime. Then the growth of vehicle queues at all
%streets is unavoidable. In order to slow down the growth of the queues and in order to
%resolve them as fast as possible after the traffic demand decreases, it is best to
%maximize the throughput \cite{Gazis2002}, considering spill-back effects. This principle
%is being followed by the stabilization strategy, which assigns longer green times
%$g_i^\max$ to the main streets having a higher service capacity $Q_i^\max$, see
%Eq.~(\ref{eq:gMax}). Still, the duration of the red-times is limited due to the fact that
%every street is served once within the desired service interval $T$. Thus, the resulting
%switching sequence tends to become periodic, as shown in the right-most bar of
%Fig.~\ref{fig:isoPeriod}.

% ------------------------------------------------------------------------------------
% ------------------------------------------------------------------------------------
\subsection{Performance at an isolated intersection}
\label{sec:SimIsolated}

Let us now compare our self-organizing control strategy with a simple fixed-time
cycle-based strategy, the cycle time of which was set to a constant value of 120 seconds.
While the switching order was set to A-B-C-D, the green times were adapted according to
the formula
\begin{equation}
    \label{eq:fixedTimeController}
    g_i^0 = \frac{u_i}{\sum_{i'} u_{i'}}
    \left( T - \sum\nolimits_{i'} \tau_{i'}^0 \right).
\end{equation}
That is, the green time $g_i^0$ of each flow $i$ was specified proportionally to the
corresponding partial utilization $u_i = Q_i / Q_i^\max$.

% ------------------------------------------------------------------------------------
\subsubsection{Constant inflows}

\begin{figure}
    \includegraphics[width=\figwidth]{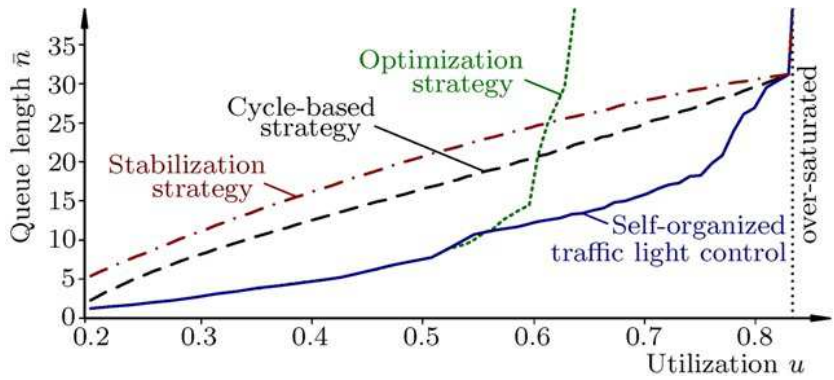}
    \caption{
Average total queue length $\bar n = \sum_i \bar n_i$ at an intersection as depicted in
Fig.~\ref{fig:singleIntersection}. The optimization strategy becomes unstable already at
medium utilizations levels $u > 0.6$ and the stabilization strategy performes always
worse than the cycle-based control. By suitably combining both inferior strategies,
however, our self-organized traffic light control performs significantly better at
\emph{all} utilization levels. This also means that the traffic flow network will enter
the over-saturated flow regime later, if at all. Therefore, traffic breakdowns during
rush hours can be avoided or at least delayed, and the recovery from congestion will
proceed faster.}
    \label{fig:isoQueue}
\end{figure}

Figure~\ref{fig:isoQueue} shows the average total queue lengths $\bar n$ for the case of
regular inflows, i.e. for identical time gaps between the arriving vehicles.
Interestingly, the optimization strategy performs better than the cycle-based approach as
long as the traffic demand is low. But it fails at high utilizations $u > 0.6$, which is
due to the strong prioritization of the main streets, where a higher throughput can be
reached over a short optimization horizon. In the course of time, the side streets are,
therefore, served too seldom or too short.

The stabilization strategy of Sec.~\ref{sec:Stab}, in contrast, is stable at all
utilization levels $u$, but it is associated with longer queues and, thus, longer average
waiting times. However, the combined strategy of Sec.~\ref{sec:Combined} starts serving
the side streets already \emph{before} their queues grow too long. For this reason, the
corresponding self-organized control strategy reaches a significant reduction of queue
lengths and waiting times at \emph{all} utilization levels.

% ------------------------------------------------------------------------------------
\subsubsection{Variable inflows}

\begin{figure}
    \includegraphics[width=\figwidth]{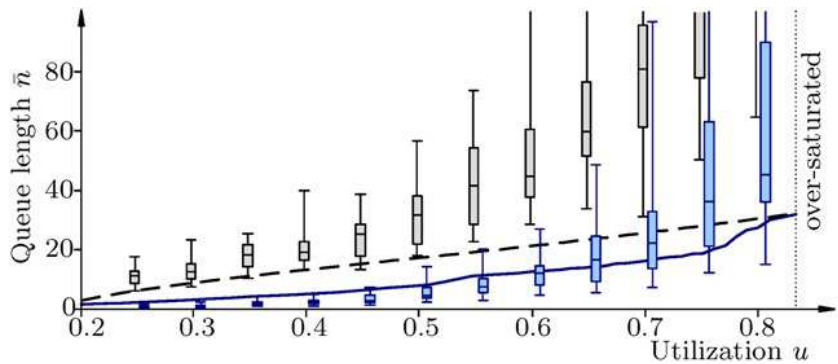}
    \caption{
Box-Whisker plot for the queue lengths at an isolated intersection (see
Fig.~\ref{fig:singleIntersection}) now with stochastic inflows. For the cycle-based
control, the queue lengths are significantly higher compared to the case with regular
inflow (see Fig.~\ref{fig:isoQueue}). In contrast, our self-organized control strategy
manages to adjust to the stochastic variations in a flexible way, which leads to a
reduction in both the mean value and the variance of the queue lengths.}
    \label{fig:isoStoch}
\end{figure}

In the following simulation we assume that the vehicles arrive in platoons, where both
the size of the platoons as well as the time gap between them are Poisson-distributed.
Fig.~\ref{fig:isoStoch} shows the Box-Whisker plot (0-25-50-75-100 percentiles) of the
stationary queue length distribution over 25 independent simulation runs each.

Because the cycle-based control strategy cannot respond to irregular inflow patterns, the
green times are sometimes too short and sometimes too long, resulting in greater delays
at all utilization levels. In contrast, the self-organizing traffic light control has a
large degree of flexibility to adjust to randomly arriving platoons. At low utilizations
$u < 0.5$, where it is possible to serve the platoons just as they arrive, there are
almost no delays. But even at higher utilizations $u \leq 0.7$, the queue lengths are
significantly smaller compared to the case with regular inflows (see
Fig.~\ref{fig:isoStoch}). Hence, our self-optimizing traffic lights could adjust well to
the fluctuations in the inflow: The irregularly arriving platoons were served by
irregular switching sequences. Altogether, this resulted in a reduction of the
variability of the queue lengths and the related waiting times.

% ------------------------------------------------------------------------------------
% ------------------------------------------------------------------------------------
\subsection{Coordination in networks}
\label{sec:SimNetwork}

% ------------------------------------------------------------------------------------
\subsubsection{Solving the Kumar-Seidman problem}

\begin{figure}
    \includegraphics[width=\figwidth]{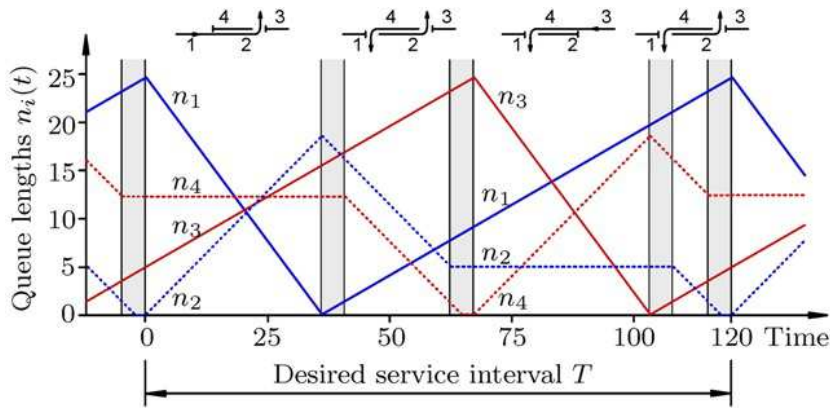}
    \caption{
Time-dependent queue lengths for the non-acyclic road network of
Fig.~\ref{fig:intersections}(b) analyzed in Fig.~\ref{fig:kumar}, but now assuming our
self-organized traffic light control. We find stable queue lengths $n_i(t)$ for the same
parameters and boundary conditions that caused a dynamic instability when a clearing
policy was applied (see Fig. \ref{fig:kumar}). The  inflows were $Q_\A = Q_\B = 1100$
vehicles per hour, the saturation flow rates were 1800 vehicles per hour and lane, and
the setup times (vertical bars) were $\tau_i^0 = 5$ seconds. Our self-organized traffic
light control stabilizes the network and serves each traffic flow once within the desired
service interval $T = 120$.}
    \label{fig:kumar_stabilized}
\end{figure}

In \ref{sec:dynamicInstabilities}, we demonstrate how a clearing policy, e.g. the
Clear-Largest-Buffer-Strategy, can behave unstable in non-acyclic networks. The same
network, illustrated in Fig.~\ref{fig:intersections}(b), shall now be operated with our
self-organized traffic light control. Figure~\ref{fig:kumar_stabilized} shows the
periodic behaviour of the queue lengths in the steady state. Our self-control succeeds to
stabilize the network, in particular because the stabilization strategy terminates
serving streets 2 and 4 as soon as the anticipated number of vehicles on street 3 or 1,
respectively, has exceeded the critical threshold $n_i^\crit(z_i)$ given by
Eq.~(\ref{eq:nCrit}). Inefficiencies due to overly long green time extensions, which were
responsible for the instability when using clearing policies, are thereby avoided.

% ------------------------------------------------------------------------------------
\subsubsection{Irregular networks}

\begin{figure}
    \includegraphics[width=\figwidth]{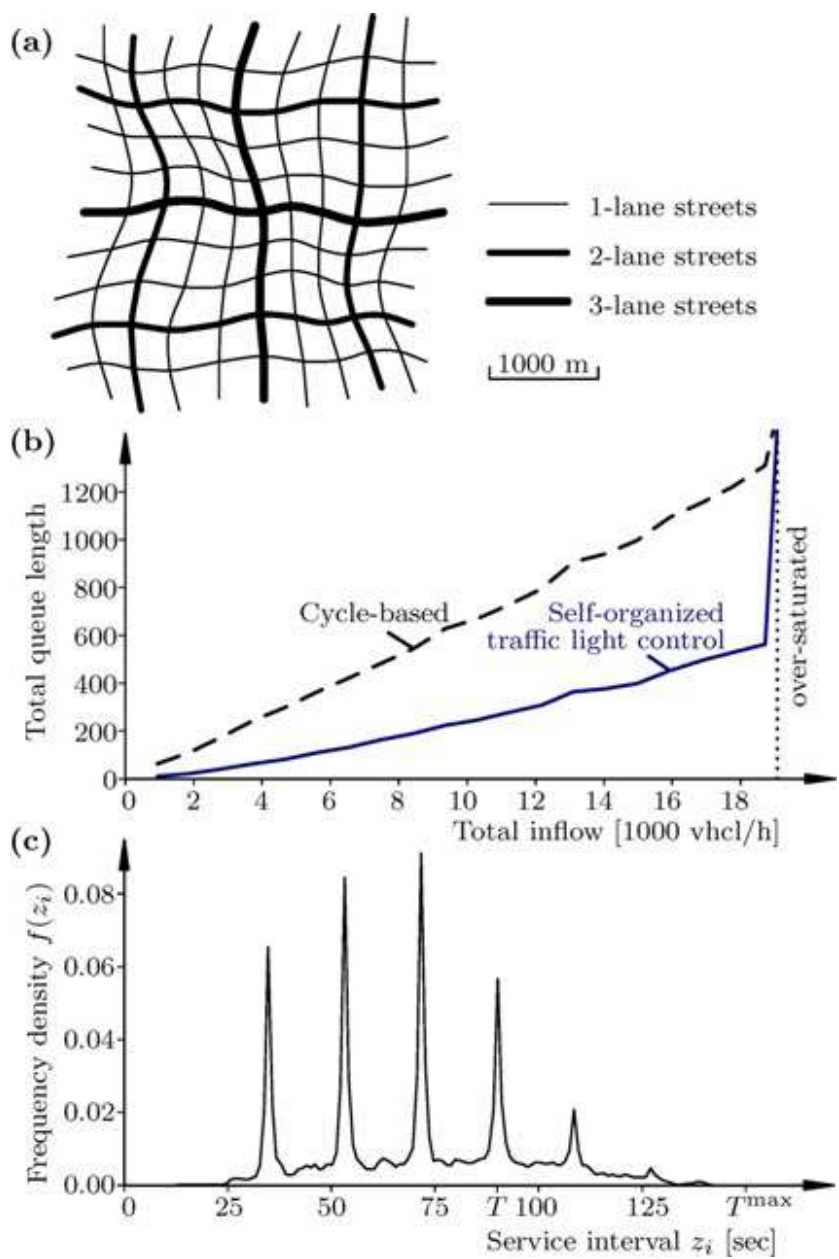}
    \caption{
(a) Road network with irregular road lengths and different numbers of lanes. (b) Average
queue lengths for different inflow rates. In contrast to the cycle-based strategy, our
approach reaches a substantial reduction in the average queue length and related travel
times. (c) The frequency density distribution $f(z_i)$ exhibits prominent peaks at
different service intervals $z_i$. This indicates a self-organized coordination with a
tendency of cycle times that are multiples of $\approx$ 18 seconds. Interestingly enough,
a cycle of 18 seconds duration does not occur itself.}
    \label{fig:irregNw}
\end{figure}

Let us now consider a $9 \times 9$ lattice road network, where both the length and the
number of lanes of the road sections are irregular. The network layout is depicted in
Fig.~\ref{fig:irregNw}(a). The saturation flow is 1800 vehicles per hour and lane, and
the speed limit is 50 km/h on all streets. Traffic enters and leaves the network at its
boundary links and distributes according to a constant turning fraction of $\alpha_{ij} =
10\%$ turning left and right, while $\alpha_{ij} = 80\%$ go straight ahead at each
intersection. The arrival rate at each entry point is proportional to the corresponding
number of lanes. For the operation of the traffic lights we assume a setup time of
$\tau_i^0 = 5$ seconds, a desired service interval of $T = 90$ seconds, and a maximum
service interval of $T^\max = 150$ seconds. For the fixed-time control strategy, with
which we compare our results, we chose a cycle time of 90 seconds, demand-adaptive green
times as specified by Eq.~(\ref{eq:fixedTimeController}), and random offsets between the
intersections.

Figure~\ref{fig:irregNw}(b) plots the total average queue length $\sum_i \bar n_i$ in the
stationary state (over one simulation hour) against the total traffic volume entering the
network. Above a maximum inflow of 18,700 vehicles per hour, where the first
intersections are over-saturated, neither strategy can prevent the queues from growing.
Up to this value, however, our self-organized control strategy exhibits significantly
smaller queue lengths in contrast to the cycle-based strategy. This is particularly due
to the fact, that our strategy has the flexibility to switch more often at less saturated
intersections. However, the traffic lights are still coordinated, and the gain in
performance is significant. It is even higher than in the case of a single intersection.

The implicit coordination of the traffic lights becomes clear in
Fig.~\ref{fig:irregNw}(c), which has been determined for a total inflow of 10,000
vehicles per hour. It shows the frequency density distribution $f(z_i)$ of the service
intervals $z_i$ over all traffic lights in the network exhibits prominent peaks at
fractions of a basic frequency. This regularity indicates that a distinct periodicity in
the switching sequence has emerged. Even though many traffic flows are served exactly
once within $T$, the period of the actual switching sequences is much smaller. This is
because some traffic flows are served several times within the time period $T$.
Therefore, it may take several intervals $T$ before a switching sequence repeats (see
also Fig.~\ref{fig:isoPeriod}). Nevertheless, the service interval does not exceed the
maximum service interval $T^\max$.

% ------------------------------------------------------------------------------------
% ------------------------------------------------------------------------------------
% ------------------------------------------------------------------------------------
\section{Conclusions and Outlook}
\label{sec:Conclusion}

In this paper, we have proposed a self-organized traffic light control based on
decentralized, local interactions. A visualization of its functional principles and
properties is provided on the webpage
\href{http://traffic.stefanlaemmer.de}{\texttt{http://traffic.stefanlaemmer.de}}, which
includes many video animations of traffic simulations. The corresponding self-control
concept is based on Eq. (\ref{eq:Combined}), together with the specifications in
Eqs.~(\ref{eq:pi}), (\ref{eq:nCrit}), and (\ref{eq:gMax}). It differs from previous
signal control approaches in the following points:
\begin{enumerate}
 \item
It reaches a superior performance by a non-periodic service, which is more flexible. A
periodic traffic light control may, nevertheless, emerge, if the street network is
grid-like and the incoming flows and turning fractions (or the boundary conditions) are
periodic.
 \item
The variation of waiting times is surprisingly small, i.e. the average waiting times are
well predictable, even though the sequence and duration of green times are basically
unpredictable.
 \item
Our simulation results suggest that a substantial reduction of the average travel times,
and therefore also of the fuel consumption and CO$_2$ emission, could be reached
\cite{Lammer2007}.
\item
The greatest gain in performance compared to previous traffic control approaches is
expected (a) for strongly varying inflows, (b) irregular road networks, (c) large
variations of the flows in different directions and among neighboring traffic lights, (c)
at night, where single vehicles should be served upon their arrival at the traffic light.
\end{enumerate}

The success principle behind the superior performance of our decentralized self-control
concept is the {\it combination} of two inferior strategies, a stabilization and an
optimizing rule, which allows for a varying sequence of traffic phases and a spatially
coordinated, non-cyclical operation. The new approach can be easily integrated into a
given traffic control environment (i.e. it is compatible with pre-specified controls at
certain intersections). The decentralized sensor, communication and control concept is
potentially less costly than centralized control concepts, and it can be set up in a way
ensuring that traffic lights are still operational when measurement sensors or
communication fail. Extensions to multi-phase operation and prioritization of public
traffic will be presented in a forthcoming paper. Similar decentralized self-control
strategies could be applied to the coordination of logistic or production processes and
even to the coordination of work-flows in companies and administrations. (In fact, it
might be easier to implement them in one of {\it these} systems, as traffic legislation
and/or operation would first have to be adjusted to allow for the more efficient,
non-cyclical traffic light operation.)

\begin{figure}
    \includegraphics[width=\figwidth]{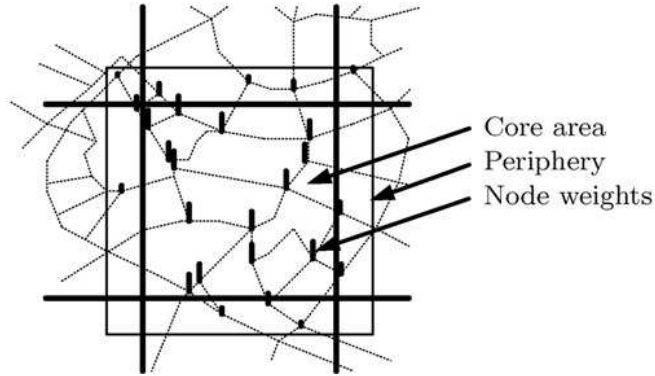}
    \caption{
Illustration of the subdivision of a road network into non-overlapping subnetworks
(``core areas'') and definition of peripheral boundary areas as proposed in
Ref.~\cite{patent}.}
    \label{fig:patent}
\end{figure}

The proposed self-organized traffic light control is the first concrete realization of
the approach suggested in a previous patent \cite{patent}. There, the road network has
been completely subdivided into non-overlapping subnetworks (``core areas''). Moreover,
each of the subnetworks was extended by additional neighboring nodes that define a
boundary area (``periphery''). The boundary areas overlap with parts of neighboring core
areas and serve the coordination between the traffic light controls of the subnetworks
(see Fig.~\ref{fig:patent}). In each core area, one first determines highly performing
solutions, assuming given traffic flows in the boundary areas. A traffic light control
for the full network is then defined by a combination of highly performing traffic light
controls for the subnetworks. The combination which performs best in the full network is
finally applied (where the best solution for the full network is not necessarily the
combination of the best solutions for the subnetworks).

The realization proposed in this paper assumes the smallest possible specification of the
subnetworks, namely the single nodes and the corresponding set of ingoing links. The
neighboring nodes constitute the respective boundary area of a node. The boundary areas
are involved in the short-term anticipation of traffic flows in the associated
subnetworks. To determine highly performing traffic light controls in the subnetworks, we
apply optimization and stabilization strategies (see Secs.~\ref{sec:Prio}
and~\ref{sec:Stab}). The traffic light control in the full network is then implemented as
follows: The stabilization strategy is applied at nodes $i$ where the set $\Omega_i$ is
non-empty, while the optimization strategy is applied at the other nodes. This traffic
light control performs better in the full network than applying the optimization strategy
at all nodes, although the latter would minimize the travel times locally. The higher
performance results from avoiding spill-back effects (see Sec.~\ref{sec:Combined}), which
would eventually block other intersections.

% ------------------------------------------------------------------------------------
% ------------------------------------------------------------------------------------
% ------------------------------------------------------------------------------------
% ACKNOWLEDGEMENT
\ack

The authors are grateful to the German Research Foundation (DFG research projects He
2789/5-1, 8-1) and the Volkswagen foundation (project I/82 697) for partial financial
support of this research.

% ------------------------------------------------------------------------------------
% ------------------------------------------------------------------------------------
% ------------------------------------------------------------------------------------
\appendix

\section{General problems of network flow coordination}
 \label{sec:Coordination}

% ------------------------------------------------------------------------------------
% ------------------------------------------------------------------------------------
\subsection{Dynamic instabilities}
\label{sec:dynamicInstabilities}

In game theory, it is known that the interaction of selfish agents can lead to
inefficiencies, such as social dilemmas \cite{Huberman1997}. Therefore, a decentralized
flow optimization by each single intersection is not necessarily optimal for the network.
In fact, it is not a successful strategy to design a control-algorithm for a single
intersection and to operate the network based on such local intersection controls. Even
if the control at each intersection minimizes the local increase of travel time, the
dynamic coupling of neighboring intersections in the network can lead to inefficiencies
due to correlations in the flow dynamics (see Fig. \ref{fig:kumar}). The problem is
either the loss of service times by frequent switches of traffic lights or the lack of
coordination between them, which may increase the average waiting times. In other words,
the intersections may not be able to handle the same amount of traffic as they could in
isolation, assuming that the arrival of vehicles is continuous. Inefficiencies reduce the
intersection capacities and cause the queues to grow longer and longer. The related
spill-back effect sooner or later blocks the flow at upstream intersections. This
phenomenon, which is referred to as dynamic instability \cite{Kumar1990}, was, for
example, demonstrated to occur under the following two conditions:
\begin{enumerate}[(i)]
    \item if the road network is non-acyclic, and
    \item if the traffic control pursues a clearing policy.
\end{enumerate}
In the following, both mechanisms shall be explained in more detail.

Regarding (i): A flow network is acyclic, if one could rank the nodes in such a way that
all flows pass the nodes from lower to higher rank. Road networks, however, can never be
acyclic. This is simply due to the fact, that there always exist paths leading from one
intersection to any other and back (not necessarily along the same route). This makes
such a ranking impossible. The critical aspect of non-acyclic networks is that
information propagates in so-called feedback loops. It means that if one intersection
sends a platoon of vehicles to one of its neighbours, it influences the time point at
which another platoon is sent back. Thus, the arrivals at an intersections are not
independent of its past switching sequence. Because these couplings do not only exist
between neighboring intersections, but between all intersections in the network, and
because these couplings have dynamically varying, travel-time related time-delays, these
feedbacks are far too complex to be anticipated locally.

Regarding (ii): Clearing policies continue serving a street until its queue has been
fully cleared \cite{Perkins1989}. They only differ in the rules selecting what street to
serve next. Such policies were shown in many experiments to be optimal at isolated
intersections \cite{Oyen1992,Lefeber2006,Eekelen2007}, but also to cause dynamic
instabilities in non-acyclic networks \cite{Kumar1990,Kumar1995}. This is for example the
case in the road network depicted in Fig.~\ref{fig:intersections}(b). Even though each
intersection alone would have been controlled optimally, as soon as they are placed next
to each other in a network, they turn out to behave unstable (see Fig.~\ref{fig:kumar}).
These facts indicate the importance to test decentralized traffic light controls in
non-acyclic networks. Unfortunately, many recently proposed approaches have been tested
either at isolated intersections or in networks with uni-directional streets only. This
may explain why most decentralized control concepts have not been practically
implemented.

Our arguments above, however, do not imply that a road network can {\em never} be
successfully controlled in a decentralized way, i.e. with independent control-algorithms
at each intersection. As shown in this paper, such a strategy is in fact possible, but it
requires to use a novel control mechanism. In Chapter~\ref{sec:TrafficLightControl}, we
have proposed such a mechanism, designed a self-organizing traffic light control, and
extended it to fulfill critical safety requirements, i.e. to comply with maximum red
times.

% ------------------------------------------------------------------------------------
% ------------------------------------------------------------------------------------
\subsection{Chaotic dynamics}
\label{sec:chaos}

Switched flow networks are known to exhibit chaotic behavior under certain conditions
\cite{Safonov2002,Wiggins2003}. In principle, this is also expected to apply to traffic
light controlled road networks. As a generic feature, chaotic behavior of a dynamical
system is characterized by an exponential divergence of initially close trajectories.
This sensitivity against small perturbations, which is a result of the intrinsic
nonlinearity of the system, is often named the ``butterfly effect'' \cite{Schuster2005}.
It may occur even without any stochasticity in the system behaviour.

Chase et al. \cite{Chase1992,Chase1993} illustrated that chaotic behavior emerges even in
very simple switched flow systems. For example, in case of a single server responsible
for serving three or more different flow directions, the resulting dynamics may be
chaotic if the server is filling one buffer up to a certain level and then switches to
another buffer (switched arrival system) \cite{Rem2003}. But also the opposite case,
where the server starts clearing a buffer as soon as its fill level exceeds a critical
threshold, exhibits chaotic behavior (switched server system with limited buffers)
\cite{Peters2004,Yu1996}. The latter case directly corresponds to a traffic light
controlled intersection with restricted queue lengths at the incoming links. The generic
mechanism leading to this behaviour can be understood by studying the the manifold
(hyperplane), in which the trajectories of the related queue lengths (reflecting buffer
fill levels) evolve. Because the underlying switching rules impose certain boundaries on
this hyperplane, the trajectories experience a so-called ``strange reflection'' if they
hit one of these boundaries. This observation allows to describe such systems in terms of
``pseudo billiard dynamics'' \cite{Blank2004,Peters2003}.

Studying the temporal evolution of vehicle positions, chaos can be observed even if the
switching sequence of the traffic lights is given. This was shown by Toledo et al.
\cite{Toledo2004} and Nagatani \cite{Nagatani2005} for a single vehicle moving through a
sequence of fixed-time controlled traffic lights. This observation is independent of
whether the distances between the traffic lights are regular or not \cite{Nagatani2006}.
In order to observe chaos, moreover, one does not even require traffic lights at all.
Wastavino et al. \cite{Wastavino2007} illustrated this for the case, where vehicles are
obstructed by yield signs.

The above examples suggest that chaotic behavior is intrinsic to vehicular flow in
traffic networks. Whereas traffic flow is statistically well predictable at an aggregate
level, it becomes highly unpredictable as soon as we want to describe its dynamics.
Predictability, however, is of great importance for the design of a traffic light control
that should be able to coordinating traffic flows, in particular to respond to large
platoons as well as single vehicles. The purpose of our proposed anticipation strategy
(see Sec.~\ref{sec:Anticipation}), therefore, is not to statistically average over the
complex nonlinear dynamics, but to cope with it on a short time scale in the most
flexible way.

% ------------------------------------------------------------------------------------
% ------------------------------------------------------------------------------------
\subsection{Limited prognosis time horizon}
\label{sec:limitedHorizon}

The unpredictable nature of traffic flow makes it particularly difficult to anticipate
traffic conditions over long time horizons. Even if we assume the streets to be equipped
with detectors and the intersections to communicate with each other, the prognosis time
horizon can hardly be larger than twice the travel time $L_i/V_i$ along the connecting
links, e.g. 30 to 40 seconds for a typical road section with $L_i \approx 300$ m and $V_i
\approx 50$ km/h. Whereas the model presented in Sec.~\ref{sec:FlowModel} can predict
well over time horizons of less than the travel time $L_i/V_i$, larger horizons obviously
need to take the switching sequence of neighboring intersections into account. If the
control decision of an intersection depends on a time horizon of more than twice the
travel time, this implies that the outcome of the control decision must be already known
to its neighbor. Such kinds of information loops are yet another complication by the
non-acyclic nature of road networks.

These considerations show that the problem of limited prognosis horizons is common to all
flexible, vehicle-responsive traffic light controls, no matter whether they are
implemented in a centralized or decentralized way. Thus, the fact that long-range
interactions are highly complex and almost impossible to predict, holds for any control
as long as it is flexibly responding to changing traffic conditions and not just imposing
a pre-defined pattern on the traffic flows, such as conventional, cycle-based controls
do. Another consequence of the limited prognosis time horizon is that any optimization is
inevitably short-sighted and, therefore, must be regarded as a potential source of
inefficiency and instability. This problem can be overcome, however, by introducing an
appropriate stabilization strategy in Sec.~\ref{sec:Stab}.

% ------------------------------------------------------------------------------------
% ------------------------------------------------------------------------------------
% ------------------------------------------------------------------------------------
% BIBLIOGRAPHY
\section*{References}

%\bibliographystyle{JHEP}
%\bibliographystyle{jphysicsB_num}
%\bibliography{references} % ACHTUNG: Update durch Aufruf von "bibtex"

\begin{thebibliography}{100}

\bibitem{Schrank2005}
D.~Schrank and T.~Lomax, {\em The 2005 urban mobility report}.
\newblock Texas Transportation Institute, 2005.

\bibitem{Helbing1997a}
D.~Helbing, {\em Verkehrsdynamik. {N}eue physikalische
  {M}odellierungskonzepte}.
\newblock Springer, Berlin, 1997.

\bibitem{Chowdhury2000}
D.~Chowdhury, L.~Santen, and A.~Schadschneider, {\it Statistical physics of
  vehicular traffic and some related systems},  {\em Phys. Rep.} {\bf 329}
  (2000), no.~4 199--329.

\bibitem{Schadschneider2002}
A.~Schadschneider, {\it Traffic flow: a statistical physics point of view},
  {\em Phys. Stat. Mech. Appl.} {\bf 313} (2002), no.~1-2 153--187.

\bibitem{Helbing2001}
D.~Helbing, {\it Traffic and related self-driven many-particle systems},  {\em
  Rev. {M}od. {P}hys.} {\bf 73} (2001) 1067--1141.

\bibitem{Helbing1995}
D.~Helbing and P.~Moln\'{a}r, {\it Social force model for pedestrian dynamics},
   {\em Phys. {R}ev. {E}} {\bf 51} (1995) 4282--4286.

\bibitem{Helbing2006c}
D.~Helbing, A.~Johansson, J.~Mathiesen, M.~H. Jensen, and A.~Hansen, {\it
  Analytical approach to continuous and intermittent bottleneck flows},  {\em
  Phys. Rev. Lett.} {\bf 97} (2006) 168001.

\bibitem{Nagatani2002}
T.~Nagatani, {\it The physics of traffic jams},  {\em Rep. Progr. Phys.} {\bf
  65} (2002) 1331--1386.

\bibitem{Kerner2004}
B.~S. Kerner, {\em The {P}hysics of {T}raffic : {E}mpirical {F}reeway {P}attern
  {F}eatures, {E}ngineering {A}pplications, and {T}heory}.
\newblock Springer, 2004.

\bibitem{Nagatani1999}
T.~Nagatani, {\it Chaotic jam and phase transition in traffic flow with
  passing},  {\em Phys. Rev. E} {\bf 60} (1999), no.~2 1535--1541.

\bibitem{Nishinari2003}
K.~Nishinari, M.~Treiber, and D.~Helbing, {\it Interpreting the wide scattering
  of synchronized traffic data by time gap statistics},  {\em Phys. Rev. E}
  {\bf 68} (2003), no.~067101.

\bibitem{Kerner2002}
B.~S. Kerner, {\it Empirical macroscopic features of spatial-temporal traffic
  patterns at highway bottlenecks},  {\em Phys. Rev. E} {\bf 65} (2002) 046138.

\bibitem{Banks1999}
J.~H. Banks, {\it Investigation of some characteristics of congested flow},
  {\em Transport. Res. Rec.} {\bf 1678} (1999) 128--134.

\bibitem{Lighthill1955}
M.~Lighthill and G.~Whitham, {\it On {K}inematic {W}aves: {II}. {A} {T}heory of
  {T}raffic {F}low on {L}ong {C}rowded {R}oads},  {\em Proc. Roy. Soc. Lond.
  Math. Phys. Sci.} {\bf 229} (1955), no.~1178 317--345.

\bibitem{Treiber2000}
M.~Treiber, A.~Hennecke, and D.~Helbing, {\it Microscopic simulation of
  congested traffic},  in {\em Traffic and {G}ranular {F}low '99: {S}ocial,
  {T}raffic, and {G}ranular {D}ynamics} (D.~Helbing, H.~J. Herrmann,
  M.~Schreckenberg, and D.~E. Wolf, eds.), pp.~365--376.
\newblock Springer, 2000.

\bibitem{Nagel1992}
K.~Nagel and M.~Schreckenberg, {\it A cellular automaton model for freeway
  traffic},  {\em J. Phys.} {\bf 2} (1992), no.~12 2221--2229.

\bibitem{Daganzo1994a}
C.~F. Daganzo, {\it The cell transmission model: a dynamic representation of
  highway traffic consistent with the hydrodynamic theory},  {\em Transport.
  Res. {B}} {\bf 28} (1994), no.~4 269--287.

\bibitem{Helbing2003a}
D.~Helbing, {\it A section-based queueing-theoretical traffic model for
  congestion and travel time analysis in networks},  {\em J. Phys. Math. Gen.}
  {\bf 36} (2003) L593--L598.

\bibitem{Daganzo1995a}
C.~F. Daganzo, {\it The cell transmission model. {II}: {N}etwork traffic},
  {\em Transport. Res. {B}} {\bf 29} (1995), no.~2 79--93.

\bibitem{Esser1997}
J.~Esser and M.~Schreckenberg, {\it Microscopic simulation of urban traffic
  based on cellular automata},  {\em Int. J. Mod. Phys. B} {\bf 8} (1997),
  no.~5 1025--1036.

\bibitem{Helbing2005}
D.~Helbing, S.~L\"ammer, and J.-P. Lebacque, {\it Self-organized control of
  irregular or perturbed network traffic},  in {\em Optimal {C}ontrol and
  {D}ynamic {G}ames} (C.~Deissenberg and R.~F. Hartl, eds.), pp.~239--274.
\newblock Springer, Dortrecht, 2005.

\bibitem{Helbing2005a}
D.~Helbing, R.~Jiang, and M.~Treiber, {\it Analytical investigation of
  oscillations in intersecting flows of pedestrian and vehicle traffic},  {\em
  Phys. Rev. E} {\bf 72} (Oct, 2005) 046130.

\bibitem{Helbing2007}
D.~Helbing, J.~Siegmeier, and S.~L\"{a}mmer, {\it Self-organized network
  flows},  {\em Networks and {H}eterogeneous {M}edia} {\bf 2} (2007), no.~2
  193--210.

\bibitem{Wastavino2007}
L.~A. Wastavino, B.~A. Toledo, J.~Rogan, R.~Zarama, V.~M. {n}oza, and J.~A.
  Valdivia, {\it Modeling traffic on crossroads},  {\em Phys. Stat. Mech.
  Appl.} {\bf 381} (2007) 411--419.

\bibitem{Gugat2005}
M.~Gugat, M.~Herty, A.~Klar, and G.Leugering, {\it Optimal {C}ontrol for
  {T}raffic {F}low {N}etworks},  {\em J. Optim. Theor. Appl.} {\bf 126} (2005),
  no.~3 589 -- 616.

\bibitem{Daganzo2007}
C.~F. Daganzo, {\it Urban gridlock: {M}acroscopic modeling and mitigation
  approaches},  {\em Transport. {R}es.{B}} {\bf 41} (2007), no.~1 49--62.

\bibitem{Zheng2007}
J.-F. Zheng, Z.-Y. Gao, and X.-M. Zhao, {\it Modeling cascading failures in
  congested traffic and transportation networks},  {\em Phys. Stat. Mech.
  Appl.} {\bf 385} (2007), no.~2 700--706.

\bibitem{Simonsen2007}
I.~Simonsen, L.~Buzna, K.~Peters, S.~Bornholdt, and D.~Helbing, {\it Stationary
  network load models underestimate vulnerability to cascading failures},  {\em
  eprint arXiv:0704.1952} (2007).

\bibitem{Brockfeld2001}
E.~Brockfeld, R.~Barlovic, A.~Schadschneider, and M.~Schreckenberg, {\it
  Optimizing traffic lights in a cellular automaton model for city traffic},
  {\em Phys. Rev. E} {\bf 64} (2001) 056132.

\bibitem{Fouladvand2001}
M.~Fouladvand and M.~Nematollahi, {\it Optimization of green-times at an
  isolated urban crossroads},  {\em Eur. Phys. J. B} {\bf 22} (2001), no.~3
  395--401.

\bibitem{Chowdhury1999}
D.~Chowdhury and A.~Schadschneider, {\it Self-organization of traffic jams in
  cities: {E}ffects of stochastic dynamics and signal periods},  {\em Phys.
  Rev. E} {\bf 59} (1999), no.~2 R1311--R1314.

\bibitem{Huang2003}
D.~Huang and W.~Huang, {\it Traffic signal synchronization},  {\em Phys. Rev.
  E} {\bf 67} (2003) 056124.

\bibitem{Sasaki2003}
M.~Sasaki and T.~Nagatani, {\it Transition and saturation of traffic flow
  controlled by traffic lights},  {\em Phys. Stat. Mech. Appl.} {\bf 325}
  (2003), no.~3-4 531--546.

\bibitem{Sekiyama2001}
K.~Sekiyama, J.~Nakanishi, I.~Takagawa, T.~Higashi, and T.~Fukuda, {\it
  Self-organizing control of urban traffic signal network},  {\em IEEE Int.
  Conf. Syst. Man. Cybern.} {\bf 4} (2001) 2481--2486.

\bibitem{Lammer2007}
S.~L\"{a}mmer, {\em Reglerentwurf zur dezentralen Online-Steuerung von
  Lichtsignalanlagen in Stra\ss{}ennetzwerken (Controller design for a
  decentralized control of traffic lights in urban road networks)}.
\newblock Ph.D. thesis. Dresden University of Technology, April, 2007.

\bibitem{Fouladvand2004}
M.~E. Fouladvand, M.~R. Shaebani, and Z.~Sadjadi, {\it Simulation of
  {I}ntelligent {C}ontrolling of {T}raffic {F}low at a {S}mall {C}ity
  {N}etwork},  {\em J.{P}hys.{S}ociety {J}apan} {\bf 73} (2004), no.~11 3209.

\bibitem{Barlovic2004}
R.~Barlovic, T.~Huisinga, A.~Schadschneider, and M.~Schreckenberg, {\it
  Adaptive traffic light control in the chsch model for city traffic},  in {\em
  Traffic and {G}ranular {F}low'03} (P.~H.~L. Bovy, S.~P. Hoogendoorn,
  M.~Schreckenberg, and D.~E. Wolf, eds.).
\newblock Springer Verlag, 2004.

\bibitem{Gershenson2005}
C.~Gershenson, {\it Self-{O}rganizing {T}raffic {L}ights},  {\em Complex
  {S}ystems} {\bf 16} (2005), no.~1 29--53.

\bibitem{Lammer2006}
S.~L\"{a}mmer, H.~Kori, K.~Peters, and D.~Helbing, {\it Decentralised control
  of material or traffic flows in networks using phase-synchronisation},  {\em
  Phys. Stat. Mech. Appl.} {\bf 363} (2006), no.~1 39--47.

\bibitem{Nakatsuji1995}
T.~Nakatsuji, S.~Seki, and T.~Kaku, {\it Development of a self-organizing
  traffic control system using neural network models},  {\em Transport. Res.
  Rec.} {\bf 1324} (1995) 137--145.

\bibitem{Ledoux1997}
C.~Ledoux, {\it An urban traffic flow model integrating neural networks},  {\em
  Transport. Res. C Emerg. Tech.} {\bf 5} (1997), no.~5 287--300.

\bibitem{Mikami1994}
S.~Mikami and Y.~Kakazu, {\it Genetic reinforcement learning for cooperative
  traffic signalcontrol},  {\em IEEE Conf. Intell.} {\bf 1} (1994) 223--228.

\bibitem{Chiu1993}
S.~Chiu and S.~Chand, {\it Self-organizing traffic control via fuzzy logic},
  {\em Decis. Contr.} {\bf 2} (1993) 1897--1902.

\bibitem{Trabia1999}
M.~B. Trabia, M.~S. Kaseko, and M.~Ande, {\it A two-stage fuzzy logic
  controller for traffic signals},  {\em Transport. Res. C Emerg. Tech.} {\bf
  7} (1999), no.~6 353--367.

\bibitem{Hoar2002}
R.~Hoar, J.~Penner, and C.~Jacob, {\it Evolutionary swarm traffic: if ant roads
  had traffic lights},  {\em Proc. Congr. Evol. Comput.} {\bf 2} (2002) 1910 --
  1915.

\bibitem{Perkins1994}
J.~R. Perkins, C.~Humes, and P.~R. Kumar, {\it Distributed {S}cheduling of
  {F}lexible {M}anufacturing {S}ystems: {S}tability and {P}erformance},  {\em
  IEEE Trans. Robot. Autom.} {\bf 10} (1994), no.~2 133--141.

\bibitem{Chase1992}
C.~Chase and P.~J. Ramadge, {\it On real-time scheduling policies for flexible
  manufacturing systems},  {\em IEEE Trans. Automat. Contr.} {\bf 37} (1992),
  no.~4 491--496.

\bibitem{Burgess1997}
K.~Burgess and K.~M. Passino, {\it Stable scheduling policies for flexible
  manufacturing systems},  {\em IEEE Trans. Automat. Contr.} {\bf 42} (1997),
  no.~3 420--425.

\bibitem{Righter2002}
R.~Righter, {\it Scheduling in {M}ulticlass {N}etworks with {D}eterministic
  {S}ervice {T}imes},  {\em Queueing {S}ystems} {\bf 41} (2002), no.~4 305 --
  319.

\bibitem{Savkin2002}
A.~V. Savkin and R.~J. Evans, {\em Hybrid {D}ynamical {S}ystems}.
\newblock Birkh\"{a}user, Boston, 2002.

\bibitem{Lefeber2004}
E.~Lefeber, {\em Nonlinear Models for Control of Manufacturing Systems},
  ch.~Nonlinear Models for Control of Manufacturing Systems, pp.~71--83.
\newblock Wiley, 2004.

\bibitem{Lan2006}
W.-M. Lan and T.~L. Olsen, {\it Multiproduct systems with both setup times and
  costs: Fluid bounds and schedules},  {\em Oper. Res.} {\bf 54} (2006), no.~3
  505--522.

\bibitem{Eekelen2007a}
J.~A. W.~M. Eekelen, {\em Modelling and control of discrete event manufacturing
  flow lines}.
\newblock PhD thesis, Eindhoven University of Technology, 2007.

\bibitem{Schutter1999}
B.~de~Schutter and B.~de~Moor, {\it The extended linear complementarity problem
  and the modeling and analysis of hybrid systems},  in {\em Hybrid {S}ystems
  {V}} (P.~Antsaklis, W.~Kohn, M.~Lemmon, A.~Nerode, and S.~Sastry, eds.),
  vol.~1567 of {\em Lecture Notes in Computer Science}, pp.~70--85.
\newblock Springer-Verlag, Berlin, 1999.

\bibitem{Schutter2002}
B.~de~Schutter, {\it Optimizing acyclic traffic signal switching sequences
  through an extended linear complementarity problem formulation},  {\em Eur.
  J. Oper. Res.} {\bf 139} (2002), no.~2 400--415.

\bibitem{Lefeber2006}
E.~Lefeber and J.~E. Rooda, {\it Controller design for switched linear systems
  with setups},  {\em Phys. Stat. Mech. Appl.} {\bf 363} (2006), no.~1 48--61.

\bibitem{Perkins1989}
J.~Perkins and P.~R. Kumar, {\it Stable, distributed, real-time scheduling of
  flexible manufacturing/assembly/diassembly systems},  {\em I{EEE} {T}rans.
  {A}utomat. {C}ontrol} {\bf 34} (1989), no.~2 139--148.

\bibitem{Kumar1995}
P.~Kumar and S.~P. Meyn, {\it Stability of queueing networks and scheduling
  policies},  {\em IEEE Trans. Automat. Contr.} {\bf 40} (1995), no.~2
  251--260.

\bibitem{Kumar1990}
P.~R. Kumar and T.~I. Seidman, {\it Dynamic instabilities and stabilization
  methods in distributed real-time scheduling of manufacturing systems},  {\em
  IEEE Trans. Automat. Contr.} {\bf 35} (1990), no.~3 289--298.

\bibitem{Humes1994}
C.~Humes, {\it A regulator stabilization technique: {K}umar-{S}eidman
  revisited},  {\em IEEE Trans. Automat. Contr.} {\bf 39} (1994), no.~1
  191--196.

\bibitem{Reiman1998}
M.~I. Reiman and L.~M. Wein, {\it Dynamic scheduling of a two-class queue with
  setups},  {\em Oper. Res.} {\bf 35} (1998), no.~4 532--547.

\bibitem{Dai2004}
J.~G. Dai and O.~B. Jennings, {\it Stabilizing {Q}ueueing {N}etworks with
  {S}etups},  {\em Math. Oper. Res.} {\bf 29} (2004), no.~4 891--922.

\bibitem{Savkin1998}
A.~V. Savkin, {\it Controllability of complex switched server queueing networks
  modelled as hybrid dynamical systems},  in {\em Decis. Contr.}, vol.~4,
  pp.~4289--4293, 1998.

\bibitem{Lammer2007a}
S.~L\"{a}mmer, R.~Donner, and D.~Helbing, {\it Anticipative control of switched
  queueing systems},  {\em Eur. Phys. J. B Condens. Matter} ({2007, in press}).

\bibitem{Wiggins2003}
S.~Wiggins, {\em Introduction to Applied Nonlinear Dynamical Systems and
  Chaos}.
\newblock Texts in Applied Mathematics. Springer, 2003.

\bibitem{Janson1991}
B.~N. Janson, {\it Dynamic traffic assignment for urban road networks},  {\em
  Transport. Res. B} {\bf 25} (1991), no.~2-3 143--161.

\bibitem{Bowman2001}
J.~L. Bowman and M.~E. Ben-Akiva, {\it Activity-based disaggregate travel
  demand model system with activity schedules},  {\em Transport. Res. Pol.
  Pract.} {\bf 35} (2001), no.~1 1--28.

\bibitem{Lan2001}
C.-J. Lan, {\it Adaptive turning flow estimation based on incomplete
  detectorinformation for advanced traffic management},  in {\em Intell.
  Transport. Syst.}, pp.~830--835, IEEE, 2001.

\bibitem{Daganzo1998}
C.~F. Daganzo, {\it Queue spillovers in transportation networks with a route
  choice},  {\em Transport. Sci.} {\bf 32} (1998), no.~1 3--11.

\bibitem{Herty2004}
M.~Herty and A.~Klar, {\it Simplified dynamics and optimization of large scale
  traffic networks},  {\em Math. Model. Meth. Appl. Sci.} {\bf 14} (2004),
  no.~4 579--601.

\bibitem{Garavello2006}
M.~Garavello and B.~Piccoli, {\it Traffic {F}low on a {R}oad {N}etwork {U}sing
  the {A}w{R}ascle {M}odel},  {\em Comm. Part. Differ. Equat.} {\bf 31} (2006),
  no.~2 243--275.

\bibitem{Troutbeck1997}
R.~J. Troutbeck and W.~Brilon, {\it Unsignalized intersection theory},  in {\em
  Traffic {F}low {T}heory: {A} {S}tate-of-the-{A}rt {R}eport} (N.~Gartner,
  H.~Mahmassani, C.~J. Messer, H.~Lieu, R.~Cunard, and A.~K. Rathi, eds.),
  pp.~8.1--8.47.
\newblock Transportation Research Board, 1997.

\bibitem{Webster1958}
F.~V. Webster, {\it Traffic {S}ignal {S}ettings},  {\em Road Research
  Laboratory Technical Paper} {\bf 39} (1958) 1--44.

\bibitem{Smith2001}
M.~Smith, J.~Clegg, and R.~Yarrow, {\it Modeling traffic signal control},  in
  {\em Handbook of Transport Systems and Traffic Control} (K.~J. Button and
  D.~A. Hensher, eds.), ch.~34, pp.~503--526.
\newblock Pergamon, 2001.

\bibitem{Arfken1995}
G.~B. Arfken and H.-J. Weber, {\em Mathematical Methods for Physicists}.
\newblock Academic Press, 4~ed., 1995.

\bibitem{Papadimitriou1999}
C.~H. Papadimitriou and J.~N. Tsitsiklis, {\it The {C}omplexity of {O}ptimal
  {Q}ueuing {N}etwork {C}ontrol},  {\em Math. Oper. Res.} {\bf 24} (1999),
  no.~2 293--305.

\bibitem{Porche1996}
I.~Porche, M.~Sampath, R.~Sengupta, Y.-L. Chen, and S.~Lafortune, {\it A
  decentralized scheme for real-time optimization of traffic signals},  in {\em
  Proc. IEEE Conf. Contr. Appl.}, pp.~582--589, 1996.

\bibitem{Papageorgiou2003}
M.~Papageorgiou, C.~Diakaki, V.~Dinopoulou, and A.~K.~Y. Wang, {\it Review of
  {R}oad {T}raffic {C}ontrol {S}trategies},  {\em Proc. IEEE} {\bf 91} (2003),
  no.~12 2043--2067.

\bibitem{McDonald1991}
M.~McDonald and N.~B. Hounsell, {\it Road traffic control: Transyt and scoot},
  in {\em Concise {E}ncyclopedia of {T}raffic \& {T}ransportation {S}ystems}
  (M.~Papageorgiou, ed.), Advances in Systems Control and Information
  Engineering, pp.~400--408.
\newblock Pergamon, 1991.

\bibitem{Helbing2001a}
D.~Helbing, P.~Moln\'{a}r, I.~Farkas, and K.~Bolay, {\it Self-organizing
  pedestrian movement},  {\em Environ. Plann. Plann. Des.} {\bf 28} (2001)
  361--383.

\bibitem{Oyen1992}
M.~P. van Oyen, D.~G. Pandelis, and D.~Teneketzis, {\it Optimality of index
  policies for stochastic scheduling with switching penalties},  {\em J. Appl.
  Probab.} {\bf 29} (1992), no.~4 957--966.

\bibitem{Rothkopf1984}
M.~H. Rothkopf and S.~A. Smith, {\it There are no undiscovered priority index
  sequencing rules for minimizing total delay costs},  {\em Oper. Res.} {\bf
  32} (1984), no.~2 451--456.

\bibitem{Serres1991}
Y.~D. Serres, {\it Simultaneous optimization of flow control and scheduling in
  a single server queue with two job classes},  {\em Oper. Res. Lett.} {\bf 10}
  (1991), no.~2 103--112.

\bibitem{Panwalkar1977}
S.~S. Panwalkar and W.~Iskander, {\it A {S}urvey of {S}cheduling {R}ules},
  {\em Oper. Res.} {\bf 25} (1977), no.~1 45--61.

\bibitem{Kumar1994}
S.~Kumar and P.~R. Kumar, {\it Performance bounds for queueing networks and
  scheduling policies},  {\em IEEE Trans. Automat. Contr.} {\bf 39} (1994),
  no.~8 1600--1611.

\bibitem{Harrison1975}
J.~M. Harrison, {\it A {P}riority {Q}ueue with {D}iscounted {L}inear {C}osts},
  {\em Oper. {R}es.} {\bf 23} (1975), no.~2 260--269.

\bibitem{Balachandran1970}
K.~R. Balachandran, {\it Parametric {P}riority {R}ules: {A}n {A}pproach to
  {O}ptimization in {P}riority {Q}ueues},  {\em Oper. {R}es.} {\bf 18} (1970),
  no.~3 526--540.

\bibitem{Duenyas1996}
I.~Duenyas and M.~P. van Oyen, {\it Heuristic {S}cheduling of {P}arallel
  {H}eterogeneous {Q}ueues with {S}et-{U}ps},  {\em Manag. Sci.} {\bf 42}
  (1996), no.~6 814--829.

\bibitem{Baras1985}
J.~S. Baras, A.~J. Dorsey, and A.~M. Makowski, {\it Two competing queues with
  linear costs and geometric service requirements: The $\mu c$-rule is often
  optimal},  {\em Adv. Appl. Probab.} {\bf 17} (1985), no.~1 186--209.

\bibitem{Gartner1975}
N.~H. Gartner, J.~D.~C. Little, and H.~Gabbay, {\it Optimization of {T}raffic
  {S}ignal {S}ettings by {M}ixed-{I}nteger {L}inear {P}rogramming {P}art {I}:
  {T}he {N}etwork {C}oordination {P}roblem},  {\em Transport. Sci.} {\bf 9}
  (November, 1975) 321--343.

\bibitem{Gartner1975a}
N.~H. Gartner, J.~D.~C. Little, and H.~Gabbay, {\it Optimization of {T}raffic
  {S}ignal {S}ettings by {M}ixed-{I}nteger {L}inear {P}rogramming {P}art {II}:
  {T}he {N}etwork {S}ynchronization {P}roblem},  {\em Transport. Sci.} {\bf 9}
  (1975), no.~4 344--363.

\bibitem{Gershenson2007}
C.~Gershenson, {\em Design and Control of Self-organizing Systems}.
\newblock PhD thesis, Center Leo Apostel for Interdisciplinary Studies, Vrije
  Universiteit Brussel, 2007.

\bibitem{Chang1992}
K.~C. Chang and D.~Sandhu, {\it Mean waiting time approximations in
  cyclic-service systems with exhaustive limited service policy},  {\em
  Perform. Eval.} {\bf 15} (1992), no.~1 21--40.

\bibitem{Chang1994}
K.~C. Chang and D.~Sandhu, {\it Delay analyses of token-passing protocols with
  limited token holding times},  {\em IEEE Trans. Comm.} {\bf 42} (1994)
  2833--2842.

\bibitem{Ram1993}
R.~Ram and N.~Viswanadham, {\it Gspn models for versatile multi-machine
  workcenters with finite buffers},  {\em IEEE Int. Conf. Syst. Man. Cybern.}
  {\bf 2} (October, 1993) 186--191.

\bibitem{Retting1997}
R.~A. Retting and M.~A. Greene, {\it Influence of traffic signal timing on
  red-light running and potential vehicle conflicts at urban intersections},
  {\em Transport. Res. Rec.} {\bf 1595} (1997) 1--7.

\bibitem{Porter2000}
B.~E. Porter and K.~J. England, {\it Predicting red-light running behavior: A
  traffic safety study in three urban settings},  {\em J. Saf. Res.} {\bf 31}
  (2000), no.~1 1--8.

\bibitem{Datta2000}
T.~K. Datta, D.~Feber, K.~Schattler, and S.~Datta, {\it Effective safety
  improvements through low-cost treatments},  {\em Transport. Res. Rec.} {\bf
  1734} (2000) 1--6.

\bibitem{Perry1999}
D.~Perry, W.~Stadje, and S.~Zacks, {\it Contributions to the theory of
  first-exit times of some compound processes in queueing theory},  {\em
  Queueing {S}ystems} {\bf 33} (1999), no.~4 369--379.

\bibitem{Abdel-Hameed2000}
M.~Abdel-Hameed, {\it Optimal control of a dam using $p^m_{\lambda,\tau}$
  policies and penalty cost when the input process is a compound poisson
  process with positive drift},  {\em J. Appl. Probab.} {\bf 37} (2000), no.~2
  408--416.

\bibitem{Savkin1998a}
A.~V. Savkin, {\it Regularizability of complex switched server queueing
  networks modelled as hybrid dynamical systems},  {\em Syst. Contr. Lett.}
  {\bf 35} (1998), no.~5 291--299.

\bibitem{Savkin2003}
A.~V. Savkin, {\it Optimal distributed real-time scheduling of flexible
  manufacturing networks modeled as hybrid dynamical systems},  in {\em
  Proceedings of the 42nd {IEEE} {C}onference on {D}ecision and {C}ontrol},
  vol.~5, pp.~5468--5471, 2003.

\bibitem{Gazis1964}
D.~C. Gazis, {\it Optimum {C}ontrol of a {S}ystem of {O}versaturated
  {I}ntersections},  {\em J. Oper. Res. Soc. Am.} {\bf 12} (1964), no.~6
  815--831.

\bibitem{Gazis2002}
D.~C. Gazis, {\em Traffic {T}heory}.
\newblock International Series in Operations Research \& Management Science.
  Springer, 2002.

\bibitem{PTV}
{\texttt{http://www.ptv.de}}.

\bibitem{patent}
D.~Helbing and S.~L\"{a}mmer, ``{V}erfahren zur {K}oordination konkurrierender
  {P}rozesse oder zur {S}teuerung des {T}ransports von mobilen {E}inheiten
  innerhalb eines {N}etzwerkes ({M}ethod for coordination of concurrent
  processes for control of the transport of mobile units within a network).''
  Patent WO/2006/122528, 2006.

\bibitem{Huberman1997}
B.~A. Huberman and R.~M. Lukose, {\it Social dilemmas and internet congestion},
   {\em Science} {\bf 277} (1997), no.~5325 535--537.

\bibitem{Eekelen2007}
J.~A. W.~M. van Eekelen, E.~Lefeber, and J.~E. Rooda, {\it State feedback
  control of switching server flowline with setups},  in {\em Proc. Am. Contr.
  Conf.}, pp.~3618--3623, 2007.

\bibitem{Safonov2002}
L.~A. Safonov, E.~Tomer, V.~V. Strygin, Y.~Ashkenazy, and S.~Havlin, {\it
  Multifractal chaotic attractors in a system of delay-differential equations
  modeling road traffic},  {\em Chaos} {\bf 12} (2002), no.~4 1006--1014.

\bibitem{Schuster2005}
H.~G. Schuster and W.~Just, {\em Deterministic Chaos}.
\newblock Wiley-VCH, 2005.

\bibitem{Chase1993}
C.~Chase, J.~Serrano, and P.~J. Ramadge, {\it Periodicity and chaos from
  switched flow systems: contrasting examples of discretely controlled
  continuous systems},  {\em IEEE Trans. Automat. Contr.} {\bf 38} (1993),
  no.~1 70--83.

\bibitem{Rem2003}
B.~Rem and D.~Armbruster, {\it Control and synchronization in switched arrival
  systems},  {\em Chaos} {\bf 13} (2003), no.~1 128--137.

\bibitem{Peters2004}
K.~Peters, J.~Worbs, U.~Parlitz, and H.-P. Wiendahl, {\it Manufacturing systems
  with restricted buffer sizes},  in {\em Nonlinear {D}ynamics of {P}roduction
  {S}ystems} (G.~Radons and R.~Neugebauer, eds.), pp.~39--54.
\newblock John Wiley \& Sons, 2004.

\bibitem{Yu1996}
G.-X. Yu and P.~Vakili, {\it Periodic and chaotic dynamics of a switched-server
  system undercorridor policies},  {\em IEEE Trans. Automat. Contr.} {\bf 41}
  (1996), no.~4 584--588.

\bibitem{Blank2004}
M.~Blank and L.~Bunimovich, {\it Switched flow systems: pseudo billiard
  dynamics},  {\em Dyn. Syst. Int. J.} {\bf 19} (2004), no.~4 359--370.

\bibitem{Peters2003}
K.~Peters and U.~Parlitz, {\it Hybrid systems forming strange billiards},  {\em
  Int. J. Bifurcat. Chaos Appl. Sci. Eng.} {\bf 19} (2003), no.~9 2575--2588.

\bibitem{Toledo2004}
B.~A. Toledo, V.~Munoz, J.~Rogan, C.~Tenreiro, and J.~A. Valdivia, {\it
  Modeling traffic through a sequence of traffic lights},  {\em Phys. Rev. E}
  {\bf 70} (2004) 016107.

\bibitem{Nagatani2005}
T.~Nagatani, {\it Chaos and dynamical transition of a single vehicle induced by
  traffic light and speedup},  {\em Phys. Stat. Mech. Appl.} {\bf 348} (2005)
  561--571.

\bibitem{Nagatani2006}
T.~Nagatani, {\it Control of vehicular traffic through a sequence of traffic
  lights positioned with disordered interval},  {\em Phys. Stat. Mech. Appl.}
  {\bf 368} (2006) 560--566.

\end{thebibliography}

\end{document}